\documentclass[twocolumn,showpacs,preprintnumbers,amsmath,amssymb]{revtex4}
\usepackage[usenames]{color}
\usepackage{graphicx}
\usepackage{dcolumn}
\usepackage{bm}
\begin{document}

\title{Effects of coarse-graining on the scaling behavior of
  long-range correlated and anti-correlated signals}

\author{Yinlin Xu$^{1,2}$, Qianli D.Y. Ma$^{3,4}$, Daniel T. Schmitt$^1$, Pedro Bernaola-Galv\'an$^5$ and Plamen~Ch.~Ivanov$^{1,3,5}\footnote{Corresponding author: plamen@buphy.bu.edu}$}
\affiliation{ $^1$Center for Polymer Studies and Department of Physics,Boston University, Boston, MA 02215, USA\\
$^2$College of Physics Science and Technology, Nanjing Normal University, Nanjing 210097, China\\
$^3$Harvard Medical School and Division of Sleep Medicine, Brigham $\&$ Women's Hospital, Boston, MA 02215, USA\\
$^4$ College of Geography and Biological Information, Nanjing University of Posts and Telecommunications, Nanjing 210003, China\\
$^5$ Departamento de F\'isica Aplicada II, Universidad de M\'alaga, 29071 M\'alaga, Spain}

\begin{abstract}

  We investigate how various coarse-graining methods affect the
  scaling properties of long-range power-law correlated and
  anti-correlated signals, quantified by the detrended fluctuation
  analysis.  Specifically, for coarse-graining in the magnitude of a
  signal, we consider (i) the Floor, (ii) the Symmetry and (iii) the
  Centro-Symmetry coarse-graining methods. We find, that for
  anti-correlated signals coarse-graining in the magnitude leads to a
  crossover to random behavior at large scales, and that with
  increasing the width of the coarse-graining partition interval
  $\Delta$ this crossover moves to intermediate and small scales. In
  contrast, the scaling of positively correlated signals is less
  affected by the coarse-graining, with no observable changes when
  $\Delta<1$, while for $\Delta>1$ a crossover appears at small
  scales and moves to intermediate and large scales with increasing
  $\Delta$. For very rough coarse-graining ($\Delta>3$) based on the
  Floor and Symmetry methods, the position of the crossover
  stabilizes, in contrast to the Centro-Symmetry method where the
  crossover continuously moves across scales and leads to a random
  behavior at all scales, thus indicating a much stronger effect of
  the Centro-Symmetry compared to the Floor and the Symmetry
  methods. For coarse-graining in time, where data points are averaged
  in non-overlapping time windows, we find that the scaling for both
  anti-correlated and positively correlated signals is practically
  preserved.  The results of our simulations are useful for the
  correct interpretation of the correlation and scaling properties of
  symbolic sequences.
\end{abstract}

\maketitle

\section{INTRODUCTION}

Certain complex physical and biological systems have no characteristic
scale and exhibit long-range power-law correlations.  Due to nonlinear
mechanisms controlling the underlying interactions, the output signals
of complex systems are also typically nonstationary, characterized by
embedded trends and heterogeneous segments with different local
statistical properties. Traditional methods such as power-spectrum and
auto-correlation analysis~\cite{non,Hurst1951,Mandelbrot1969} are not
suitable for nonstationary signals.  To address this problem,
detrended fluctuation analysis (DFA) was developed to more accurately
quantify long-range power-law correlations embedded in a nonstationary
time series \cite{CKDFA1,taqqu95}.  In addition to quantifying
scale-invariant features, DFA has been used to also detect
characteristic scales in non-homogeneous
signals~\cite{carpena07,vis97}.

The DFA method has been successfully applied to quantify the output
dynamics of various physical and biological systems including
meteorology~\cite{Ivanova2003}, climate temperature
fluctuations~\cite{Bundeatm,bundetem,talknertem2000}, river flow and
discharge~\cite{Montanari2000,Matsoukas2000},
economics~\cite{vandewallepre1998,Liu99,janosiecopha1999,Ivanov-eco-04},
neural receptors in biological systems~\cite{bahareuph2001},
DNA~\cite{CKDFA1,SVDFA1,SMDFA1,mantegnaprl1994,mantegnaprl1996,Buldyrev,Pedro-1996},
cardiac
dynamics~\cite{crossoverCK,HOcirc1997,plamenuropl1999,Pikkujamsaheartcir1999,makikallioheartamjcardiol1999,toweillheartmed2000,bundesleep2000,Laitio2000,Yosef2001,plamenchaos2001,Yosef2003,Ivanov-IEEE-07,Ivanov-PNAS-07,Ivanov-AJP-07,Ivanov-IEEE-09},
and human gait~\cite{hos,Ashkenazy02,Ivanov-PRE-2009}.  DFA has been
also utilized to identify transitions across different states of the
same system characterized by different scaling behavior, e.g., the
scaling exponent for heart-beat intervals discriminates between
healthy and sick individuals~\cite{crossoverCK}, wake and sleep
state~\cite{plamenuropl1999,bundesleep2000}, and across different
sleep stages and circadian phases~\cite{Ivanov-IEEE-07}.  Further, it
has been shown that the DFA scaling exponent obtained from symbolic
DNA sequences relates to the evolutionary degree of various
organisms~\cite{SVDFA1}.

To understand the intrinsic dynamics of a given system, it is
important to know how its intrinsic nature (e.g. nonstationarities) or
external manipulations such as data pre-processing can affect the
results. In previous studies we have investigated the effect of
various data artifacts on the DFA scaling analysis of long-range
power-law correlated signals.  Specifically, we considered different
types of nonstationarities associated with different trends present in
the signal, e.g., polynomial, sinusoidal and power-law
trends~\cite{kun} Further, we have studied effects of
nonstationarities that are often encountered in real data or result
from ``standard'' data pre-processing approaches, e.g., signals with
segments removed, signals with random spikes as well as signals with
different local behavior \cite{Zhichen_PRE_2002}. We have also
investigated the effects of linear and nonlinear filtering of
signals~\cite{chen05} and the effects of extreme data
loss~\cite{Ma09}. Comparative studies of the performance of the DFA
method and other scaling analysis methods are presented
in~\cite{taqqu95,Xu05,Amir07}.

In this paper we focus on the effect of different coarse-graining
approaches on the scaling properties of correlated signals quantified
by the DFA.

{\it Coarse-graining in the magnitude of a signal.} It consists of the
discretization of data and is frequently imposed by the nature of the
measurements, i.e., the limitations on accuracy of instruments,
acquisition-data rate or even data-storage requirements. In addition,
while in many situations coarse-graining is imposed by the intrinsic
nature of the data, in other cases coarse-graining is applied at a
latter stage in order, for example, to compute certain information
theory measures derived from Shannon entropy. These functionals can be
applied only to symbolic or binned time series because they take as
arguments probability distributions (in practice, we never have access
to the full distribution function but only to a binned version of it).

One of these functionals, especially suitable when dealing with
correlations in time series, is the mutual information. It is well
known that mutual information is closely related to the correlation
function but, contrary to it, mutual information captures the complete
dependence structure, including nonlinear correlations
\cite{Li89,He95}. The estimators of mutual information require the
data to be binned, i.e., the range of the data is partitioned and each
element of the time series is assigned to an interval of the
partition.  This transformation is equivalent to the coarse-graining
procedures described in this study, and depending on the details of
such transformation the correlation structure of the signal can be
modified. For example, it is known that uniform partitions in some
situations can lead to misleading results \cite{Ce05}. The measure of
correlations using mutual information has been used in several fields:
DNA sequences \cite{Ho03}, physiological signals \cite{Qu02}, quantum
information \cite{Gr05}, complex systems \cite{Ma92,Wi07}.

{\it Coarse graining in time (or space)}. 
It consists of substituting the values of the signal in an interval by
its mean value. Again, this procedure may appear as a direct effect of
the measurement (e.g. sonometers or pyranometers integrate the input
during the measurement interval) or due to further modification of
data. It is common to use this coarse-graining to smooth out the short
scale heterogeneities of a signal. For example it has been recently
used to analyze the long-scale structure of
DNA~\cite{Pedro-1999,Oliver01,Oliver02,Pedro-2003}. Another typical
application of coarse-graining in time is the mean-field approximation
where one considers the mean effect of the interactions during a given
period (in time or space) in order to simplify the problem. For
example, such techniques have been used to study the growth of fractal
structures in the framework of mean-field
approximation~\cite{Postnikov07}.

The outline of this paper is as follows. In Sec.~\ref{method}, we
review the Fourier filtering method for generating long-range
power-law correlated signals and the DFA method for scaling analysis,
and we introduce the Floor, Symmetry and Centro-Symmetry methods of
coarse-graining in the magnitude of a signal and the coarse-graining
in time method. In Sec.~\ref{Results} we compare the scaling
properties before and after coarse-graining for both anti-correlated
and positively correlated signals to study the effects of various
coarse-graining approaches on the DFA scaling. In Sec.~\ref{Summary}
we summarize our findings.

\section{Method}\label{method}

\subsection{Fourier filtering method}\label{Fourier filtering method}
Using a modified Fourier filtering method \cite{MFFM}, we generate
stationary uncorrelated, correlated, and anti-correlated signals
$x(i)$ ($i=1,2,3,...,N_{\mbox{\scriptsize max}}$) with a zero mean and
standard deviation $\sigma=1$. This method consists of the following
steps:

(a) First, we generate a random uncorrelated and Gaussian distributed
sequence {$\eta(i)$} and calculate the Fourier transform coefficients
{$\eta(q)$}.

(b) The desired signal $x(i)$ must exhibit correlations, which are
defined by the form of the power spectrum
\begin{equation}
S(q)=\langle x(q)x(-q) \rangle \sim q^{-\beta}, \label{M1}
\end{equation}
where {$x(q$)} are the Fourier transform coefficients of
{$x(i)$},$\beta$ is the Fourier spectrum exponent. Thus, we generate
{$x(q$)} using the following transformation:
\begin{equation}
x(q)=[S(q)]^{1/2}\eta(q) \sim \frac{\eta(q)}{q^{\beta/2}}, \label{M2}
\end{equation}
where $S(q)$ is the desired power spectrum in Eq.~(\ref{M1}).

(c) We calculate the inverse Fourier transform of {$x(q$)} to obtain
$x(i)$.

Thus, the signal $x(i)$ we generate is a stationary long-range
power-law correlated signal characterized by an auto-correlation
function of the type

\begin{equation}
    C(n)\equiv\langle x_i x_{i+n} \rangle_{_i}\sim n^{-\gamma},
\label{M3}
\end{equation}
where $\gamma=1-\beta$ is the correlation exponent \cite{SVDFA1}

For all cases of coarse-graining we investigate in our study, we
always consider signals $x(i),i=1,...,N_{\rm max}$ of a fixed length
$N_{\rm max}=2^{17}$.

\subsection{Coarse-graining methods}
\label{Coarse-graining magnitude methods}
In practical applications coarse-graining is traditionally
applied to the {\it magnitude} of a signal $x(i)$, i.e.,
coarse-graining (discretizing) the values which $x(i)$ can take.
Alternatively, one can coarse-grain a signal {\it in time}, i.e.,
averaging the values $x(i+1),...,x(i+\Delta)$ of the signal within
non-overlapping time intervals of size $\Delta$. In this study we
consider the most frequently used methods of coarse-graining:

\begin{enumerate}

\item {\it Coarse-graining in the magnitude} of a signal $x(i)$: (i)
Floor method, (ii) Symmetry method, and (iii) Centro-Symmetry
method.

\item  {\it Coarse-graining in time} of a signal $x(i)$.

\end{enumerate}

\subsubsection{Coarse-graining in the magnitude}
(i) {\it Floor method}. First, we define the width of the discretization
partition interval:
\begin{equation}
\Delta=K \cdot \sigma,
\end{equation}
where $\sigma$ is the standard deviation of the signal $x(i)$, and the
$K$ is a real positive number. Next, the range of $x(i)$ values is
partitioned into consecutive non-overlapping intervals of size
$\Delta$, starting from $0$ in direction of both positive and negative
values of $x(i)$ (see schematic illustration in
Fig.\ref{fig:illustration_of Floor Method}).

\begin{figure}
    \includegraphics[width=1\columnwidth]{Fig1a}\vspace*{0.2cm}
    \includegraphics[width=0.8\columnwidth]{Fig1b}
    \caption
    { Schematic illustration of the coarse-graining in magnitude
      procedure based on the {\it Floor method}
      (Eq.~\ref{floor-1}). (a)~Coarse-graining the values of an
      original signal $x(i)$ ($\circ$) with a partition factor
      $\Delta$ to obtain a coarse-grained symbolic signal
      $\tilde{x}(i)$ ($\blacktriangle$).  $K$ is the partition
      coefficient and $\sigma$ is the standard deviation of
      $x(i)$. Arrows represent the shift from real values $x(i)$ to
      integer values (symbols) $\tilde{x}(i)$, as a result of the
      Floor coarse-graining. Horizontal dashed lines represent the
      boundaries of the discretization partition intervals.
      (b)~Illustration of the transition from real values $x(i)$ to
      integer values (symbols) for the coarse-grained signal
      $\tilde{x}(i)$. For $x(i)<0$ the values of $\tilde{x}(i)$
      correspond to the index $j$ of each partition interval
      $-j\Delta$,...,$-3\Delta$,$-2\Delta$,$-1\Delta$.  In contrast,
      for $x(i)>0$ the values of $\tilde{x}(i)$ correspond to $(j-1)$
      where $j$ is the index of each partition interval $1\Delta$,
      $2\Delta$, $3\Delta$,...,$j\Delta$. Note, that $\tilde{x}(i)=0$
      for all values of $x(i)$ in the interval $[0,1\Delta$)(see
      Eq.~\ref{floor-1}). Solid horizontal lines in (b) represent the
      intervals where all $x(i)$ values become coarse-grained into
      integers (symbols) $\tilde{x}(i)$. Open ($\circ$) and closed
      ($\bullet$) circles indicate opened- and closed-end of the
      partition intervals for $x(i)$ in the Floor method
      transformation, e.g., all values $x(i)\in[1\Delta,2\Delta)$ are
      transformed into $\tilde{x}(i)=1$; all $x(i)\in[0,1\Delta)$
      become $\tilde{x}(i)=0$, and all values $x(i)\in[-1\Delta,0)$
      become $\tilde{x}(i)=-1$ etc.
     \label{fig:illustration_of Floor Method}
     }

\end{figure}

\begin{figure*}
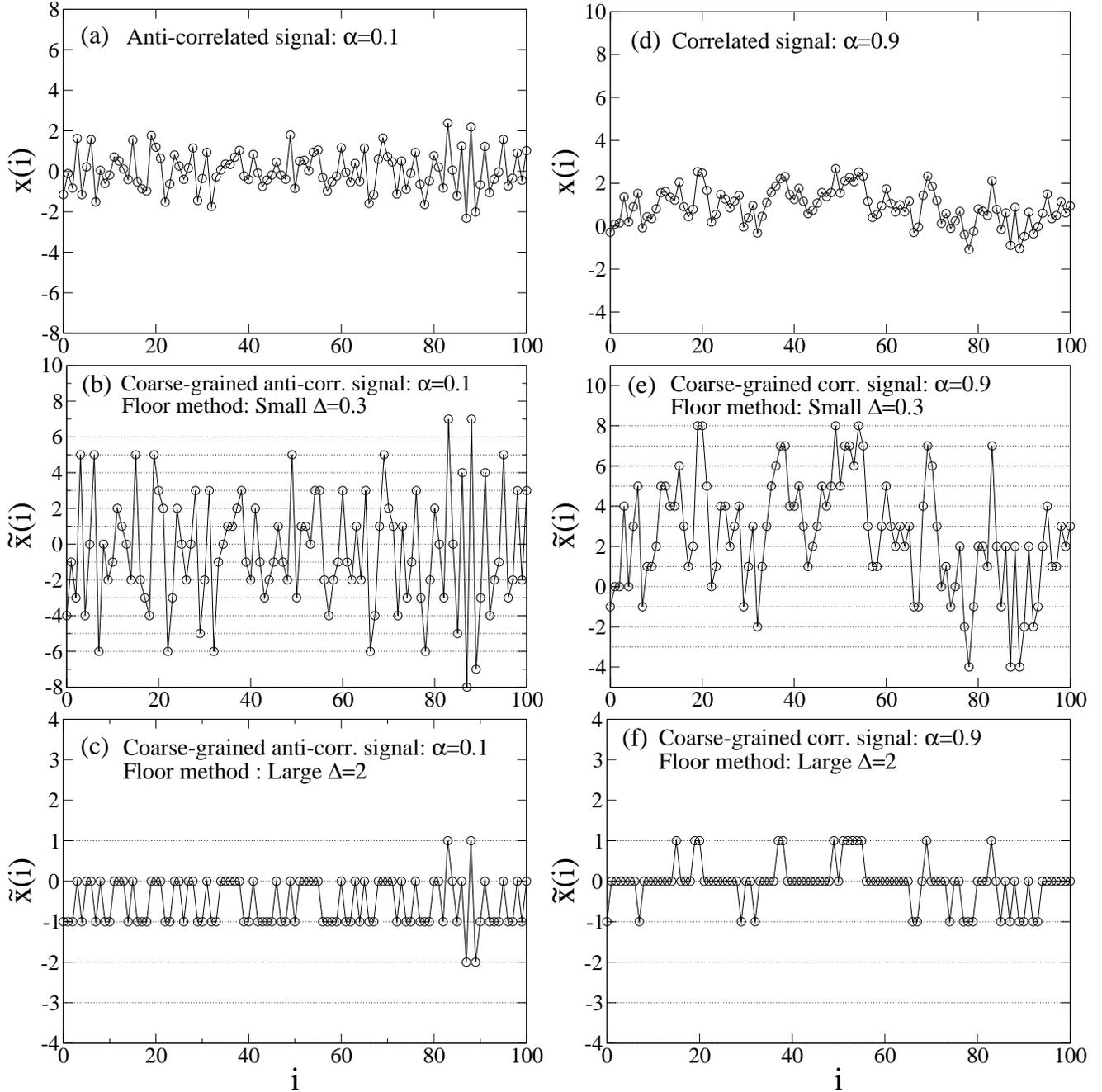

    \includegraphics[width=1\columnwidth]{Fig2a}\vspace*{0cm}
    \includegraphics[width=1\columnwidth]{Fig2d}\vspace*{0cm}
    \includegraphics[width=1\columnwidth]{Fig2b}\vspace*{0cm}
    \includegraphics[width=1\columnwidth]{Fig2e}\vspace*{0cm}
    \includegraphics[width=1\columnwidth]{Fig2c}\vspace*{0.4cm}
    \includegraphics[width=1\columnwidth]{Fig2f}\vspace*{0.4cm}
    \caption { Demonstration of the effect of the {\it Floor
        coarse-graining method} on (a)~power-law strongly
      anti-correlated signal $x(i)$ characterized by a DFA scaling
      exponent $\alpha=0.1$ (see Sec. \ref{DFA-method}) and
      (d)~power-law strongly correlated signal $x(i)$ with a DFA
      scaling exponent $\alpha=0.9$. Both signals in (a) and (d) have
      standard deviation $\sigma=1$. For small values of the width of
      the discretization partition interval, $\Delta<1$, the Floor
      coarse-graining method leads to expansion in the magnitude of
      the coarse-grained (symbolic) signal $\tilde{x}(i)$ along the
      y-axis with much larger standard deviation $\tilde{\sigma}$
      compared to $\sigma$, shown in (b) and (e).  For very large
      values of the width of the discretization partition interval,
      $\Delta\gg1$, the Floor coarse-graining practically leads to
      binary sequences for both correlated and anti-correlated
      signals, shown in (c) and (f).  We note, that for $\Delta\gg1$
      the standard deviation $\tilde{\sigma}$ of the coarse-grained
      signal $\tilde{x}(i)$ becomes smaller than $\sigma$ of the
      original signal $x(i)$ (see also Fig.\ref{fig:Effect of Floor
        method coarse-graining on DFA} a,c).  }
     \label{fig:illustration_of Floor coarse-grained signal}
\end{figure*}

All data points in the entire signal $x(i)$, which fall into the
{\it same} discretization partition interval, are then replaced by a symbol
$\tilde{x}(i)$ corresponding to the index of the partition
interval. The index of the partition interval is a positive or
negative integer (Fig \ref{fig:illustration_of Floor Method}).

In  the case of the {\it Floor coarse-graining method}, we apply 
the {\it floor rule} in order to obtain the
coarse-grained symbolic sequence $\tilde{x}(i)(i=1,...,N_{\rm max})$:

\begin{equation}
         \tilde{x}(i)\equiv \left \lfloor\frac{x(i)}{\Delta} \right \rfloor=
                           \left \lfloor\frac{x(i)}{K\cdot\sigma}\right \rfloor,
         \label{floor-1}
\end{equation}
where $\lfloor...\rfloor$ denotes the {floor} function, which for any
real value of the argument assigns the largest integer value less or
equal to the argument.  Thus, based on the floor rule we have:

\begin{itemize}

\item[(a)] for positive $x(i)$ --- any value $x(i)\in[0,\Delta)$
becomes $\tilde{x}(i)=0$; any value $x(i)\in[\Delta,2\Delta)$
becomes $\tilde{x}(i)=1$; any value $x(i)\in[2\Delta,3\Delta)$
becomes $\tilde{x}(i)=2$, etc.

\item[(b)] for negative $x(i)$ --- any value $x(i)\in[-\Delta,0)$
becomes $\tilde{x}(i)=-1$; any value $x(i)\in[-2\Delta,-\Delta)$
becomes $\tilde{x}(i)=-2$, etc.

\end{itemize}

In Fig.\ref{fig:illustration_of Floor Method} we schematically
demonstrate the coarse-graining in the magnitude procedure based
on the Floor method.

In Fig.\ref{fig:illustration_of Floor coarse-grained signal} we
illustrate the effect of coarse-graining in the magnitude based on the
Floor method for segments of two representative signals $x(i)$: one
with long-range power-law correlations and one with power-law
anti-correlations. Note, that for values of the width of the partition
interval $\Delta<1$ the Floor coarse-graining leads to an expansion of
the signals $\tilde{x}(i)$ along the y-axis. For $\Delta>1$ the Floor
coarse-graining effectively contracts the magnitude of the signals
$\tilde{x}(i)$, leading to practically binary sequences
(Fig.\ref{fig:illustration_of Floor coarse-grained signal}). This
effect appears stronger for anti-correlated signals compared to
correlated signals (compare Fig.\ref{fig:illustration_of Floor
  coarse-grained signal}c and Fig.\ref{fig:illustration_of Floor
  coarse-grained signal}f).

\begin{figure}
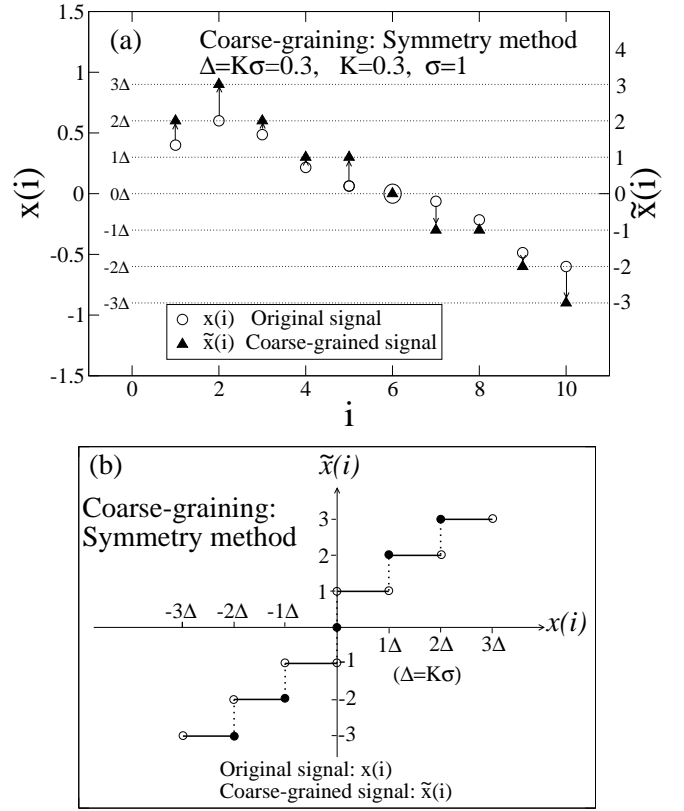

    \includegraphics[width=1\columnwidth]{Fig3a}\vspace*{0.2cm}
    \includegraphics[width=0.8\columnwidth]{Fig3b}
    \caption
    { Schematic illustration of the coarse-graining in magnitude
      procedure based on the {\it Symmetry method}
      (Eq.~\ref{symmetry-1}).  (a)~Coarse-graining the values of an
      original signal $x(i)$ ($\circ$) with a partition factor
      $\Delta$ to obtain a coarse-grained symbolic signal
      $\tilde{x}(i)$($\blacktriangle$). $K$ is the partition
      coefficient and $\sigma$ is the standard deviation of
      $x(i)$. Arrows represent the shift from real values $x(i)$ to
      integer values (symbols) $\tilde{x}(i)$, as a result of the
      Symmetry coarse-graining. Horizontal dashed lines represent the
      boundaries of the discretization intervals.  (b)~Illustration of
      the transition from real values $x(i)$ to integer values
      (symbols) for the coarse-grained signal $\tilde{x}(i)$. Note,
      that except for $\tilde{x}(i)=0$, the values of $\tilde{x}(i)$
      correspond to the index $j$ of each partition interval
      $-j\Delta$,...,$-3\Delta$,$-2\Delta$,$-1\Delta$ and
      $1\Delta$,$2\Delta$,$3\Delta$,...,$j\Delta$ of the original
      signal $x(i)$ (\ref{symmetry-1}).  For every $x(i)=0$, after the
      Symmetry coarse-graining, we obtain $\tilde{x}(i)=0$.  Solid
      horizontal lines in (b) represent the intervals of values for
      $x(i)$ which become coarse-grained into integers (symbols) for
      $\tilde{x}(i)$. Open ($\circ$) and closed ($\bullet$) circles
      indicate opened- and closed-end of the partition intervals for
      $x(i)$ in the Symmetry method transformation, e.g., all values
      of $x(i)\in[1\Delta,2\Delta)$ are transformed into
      $\tilde{x}(i)=2$, all values of $x(i)\in(0,1\Delta)$ become
      $\tilde{x}(i)=1$, all values of $x(i)\in(-1\Delta,0)$ become
      $\tilde{x}(i)=-1$, etc.
     }
     \label{fig:illustration_of Symmetry Method}

\end{figure}

(ii) {\it Symmetry method}. First, we define the width $\Delta$ of the
discretization partition interval. Next, we partition the range of
$x(i)$ values into consecutive non-overlapping partitions intervals of
size $\Delta$ starting from $0$ in direction of both negative and
positive values of $x(i)$ (see Fig.\ref{fig:illustration_of Symmetry
  Method}). Then, all data points in the entire signal $x(i)$, which
fall into the {\it same} partition interval, are replaced by the
symbol $\tilde{x}(i)$ (an integer positive or negative value)
identical to the index of the partition interval
(Fig.\ref{fig:illustration_of Symmetry Method}).

In the case of the Symmetry method, we apply the {\it symmetry rule} to obtain 
the coarse-grained symbolic sequence $\tilde{x}(i)(i=1,...,N_{\rm max})$:

\begin{equation}
  \tilde{x}(i)\equiv  {\rm round}\left [\frac{x(i)}{\Delta}+\frac{1}{2}{\rm sgn}(x(i))\right ],
         \label{symmetry-1}
\end{equation}
where

\begin{displaymath}
  {\rm sgn}(x(i))= \left\{\begin{array}{ll}
      ~1 & \textrm{if $x(i)>0$}\\
      ~0 & \textrm{if $x(i)=0  ~~~.$}\\
      -1 & \textrm{if $x(i)<0$}
\end{array} \right.
\end{displaymath}

The {round}$[...]$ function in (\ref{symmetry-1}) assigns for each
real argument the integer number closest to the argument.  Thus based
on the {\it symmetry rule} we have:

\begin{itemize}

\item[(a)] for positive $x(i)$ --- any value $x(i)\in(0,\Delta)$
  becomes $\tilde{x}(i)=1$; any value $x(i)\in[\Delta,2\Delta)$
  becomes $\tilde{x}(i)=2$; any value $x(i)\in[2\Delta,3\Delta)$
  becomes $\tilde{x}(i)=3$, etc.

\item[(b)] for negative $x(i)$ --- any value $x(i)\in(-\Delta,0)$
  becomes $\tilde{x}(i)=-1$; any value $x(i)\in(-2\Delta,-\Delta]$
  becomes $\tilde{x}(i)=-2$, etc.

\item[(c)] for $x(i)=0$ we have $\tilde{x}(i)=0$

\end{itemize}

In Fig.\ref{fig:illustration_of Symmetry Method} we schematically
demonstrate the coarse-graining in the magnitude procedure 
based on the Symmetry method.

\begin{figure*}
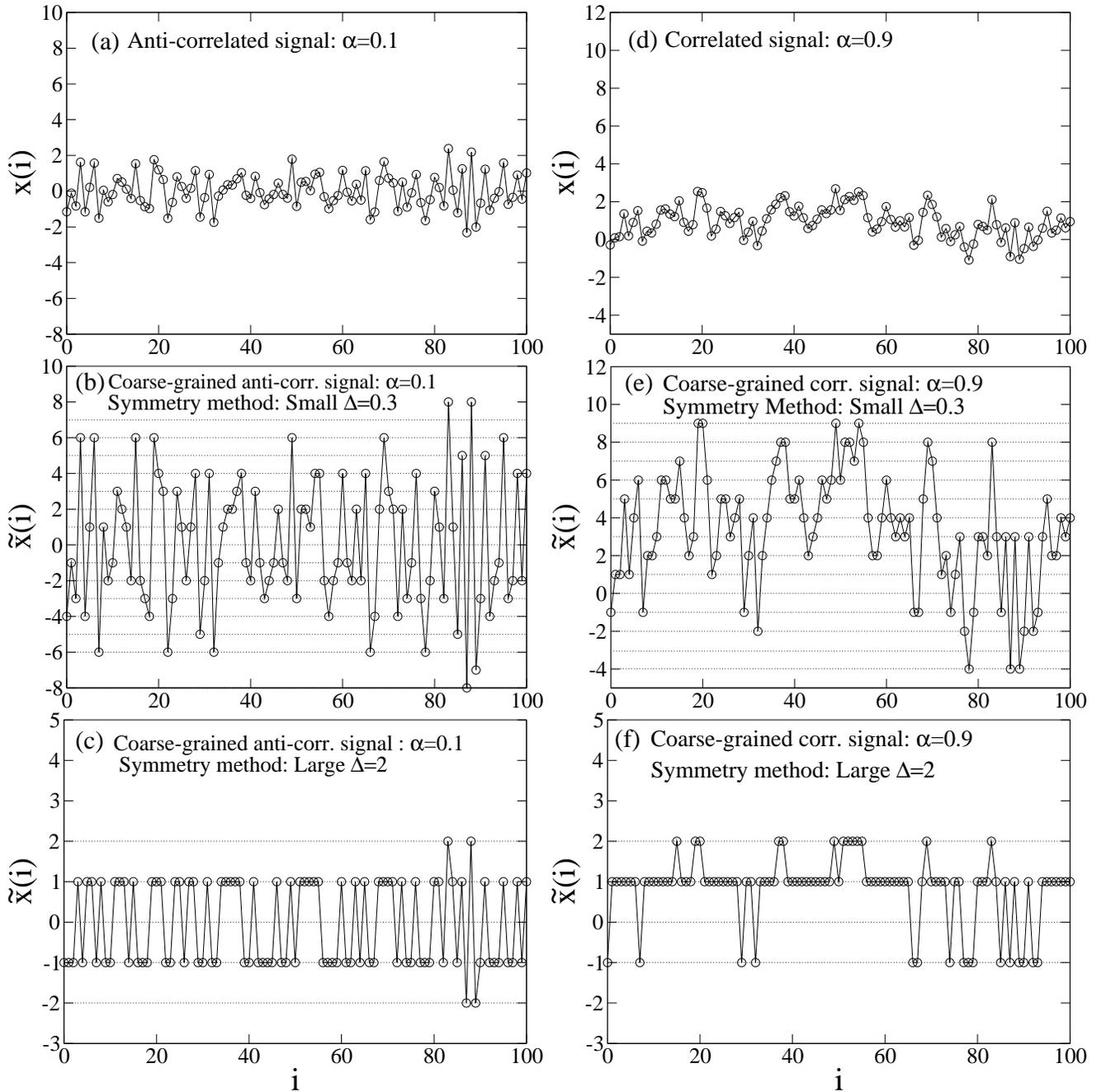

    \includegraphics[width=1\columnwidth]{Fig4a}\vspace*{0cm}
    \includegraphics[width=1\columnwidth]{Fig4d}\vspace*{0cm}
    \includegraphics[width=1\columnwidth]{Fig4b}\vspace*{0cm}
    \includegraphics[width=1\columnwidth]{Fig4e}\vspace*{0cm}
    \includegraphics[width=1\columnwidth]{Fig4c}\vspace*{0.4cm}
    \includegraphics[width=1\columnwidth]{Fig4f}\vspace*{0.4cm}
    \caption { Demonstration of the effect of the {\it Symmetry
        coarse-graining method} on (a)~power-law strongly
      anti-correlated signal $x(i)$ with a DFA scaling exponent
      $\alpha=0.1$ (see Sec. \ref{DFA-method}) and (d)~power-law
      strongly correlated signal $x(i)$ with a DFA scaling exponent
      $\alpha=0.9$. Both signals in (a) and (d) are identical to those
      shown in Fig.\ref{fig:illustration_of Floor coarse-grained
        signal}a,d, and have standard deviation $\sigma=1$.  For small
      values of the width of the discretization interval, $\Delta<1$,
      the Symmetry coarse-graining leads to expansion in magnitude of
      the coarse-grained symbolic signal $\tilde{x}(i)$ along the
      y-axis with much larger standard deviation $\tilde{\sigma}$
      compared to $\sigma$, shown in (b) and (e).  For very large
      values of the width of the partition interval, $\Delta\gg1$, the
      Symmetry coarse-graining practically leads to binary sequences
      for both correlated and anti-correlated signals, shown in (c)
      and (f).  We note, that for $\Delta<1$ the expansion in
      magnitude of the coarse-grained signal $\tilde{x}(i)$ along the
      y-axis is more pronounced for the positive values of
      $\tilde{x}(i)$ in the Symmetry method compared to the Floor
      method, while for the negative values of $\tilde{x}(i)$ the
      magnitude expansion effect along the y-axis is the same for both
      Symmetry and Floor methods (compare to
      Fig.\ref{fig:illustration_of Floor coarse-grained
        signal}b,e). We also note, that for $\Delta\gg1$, the standard
      deviation $\tilde{\sigma}$ of the Symmetry coarse-grained signal
      $\tilde{x}(i)$ is not significantly smaller than $\sigma$ of the
      original signal $x(i)$ (see also Fig.\ref{fig:Effect of
        Symmetric method coarse-graining on DFA} a,c). In contrast,
      for the Floor coarse-graining method and for $\Delta\gg1$,
      $\tilde{\sigma}$ of the coarse-grained signal $\tilde{x}(i)$ is
      significantly smaller than $\sigma$ of the original signal
      $x(i)$ for both correlated and anti-correlated signals (compare
      Fig.\ref{fig:Effect of Floor method coarse-graining on DFA} a,c
      to Fig.\ref{fig:Effect of Symmetric method coarse-graining on
        DFA} a,c).  }
     \label{fig:illustration of Symmetry coarse-grained signal}
\end{figure*}

In Fig.\ref{fig:illustration of Symmetry coarse-grained signal} we
illustrate the effect of the coarse-graining in the magnitude of a
signal based on the Symmetry method for segments of two representative
signals $x(i)$: one with long-range power-law correlations and one
with power-law anti-correlations. For comparison, in
Fig.\ref{fig:illustration of Symmetry coarse-grained signal} we show
the same identical signal segments as shown in
Fig.\ref{fig:illustration_of Floor coarse-grained signal}. For values
of the width of the partition interval $\Delta<1$ the Symmetry
coarse-graining method leads to an expansion of the signals
$\tilde{x}(i)$ along the y-axis. For signal values $x(i)<0$ the
expanding effect of the Symmetry method is identical to the effect of
the Floor method. However, for $x(i)>0$ the expanding effect of the
Symmetry coarse-graining method is stronger compared to the Floor
method. For $\Delta>1$ the Symmetry coarse-graining method effectively
contracts the magnitude of the signals $x(i)$ leading to practically
binary sequences.  (see Fig.\ref{fig:illustration of Symmetry
  coarse-grained signal}c and Fig.\ref{fig:illustration of Symmetry
  coarse-grained signal}f).  This effect is similar to the effect of
the Floor method (compare with Fig.\ref{fig:illustration_of Floor
  coarse-grained signal}c and Fig.\ref{fig:illustration_of Floor
  coarse-grained signal}f).

(iii) {\it Centro-Symmetry method}. In contrast to the Floor and
Symmetry coarse-graining methods, where the consecutive partition
intervals start from $0$, in the Centro-Symmetry method the first
partition interval is centered at $0$, i.e., $[-0.5\Delta,0.5\Delta)$,
with consecutive partitions in the positive and negative direction
starting from $0.5\Delta$ and $-0.5\Delta$ respectively.

In the case of Centro-Symmetry method we apply the {\it
  centro-symmetry} rule to obtain the coarse-grained symbolic sequence
$\tilde{x}(i)(i=1,...,N_{\rm max})$:
\begin{equation}
  \tilde{x}(i)\equiv \left \lfloor\frac{x(i)+0.5\Delta}{\Delta}\right \rfloor,
  \label{centro-symmetry-1}
\end{equation}
where $\lfloor...\rfloor$ denotes the floor function (as in
Eq.\ref{floor-1})

Based on the Centro-symmetry rule we have:

\begin{itemize}

\item[(a)] any value $x(i)\in[-0.5\Delta,+0.5\Delta)$ becomes
  $\tilde{x}(i)=0$;

\item[(b)] for positive $x(i)\geq0.5\Delta$ --- any value
  $x(i)\in[0.5\Delta,1.5\Delta)$ becomes $\tilde{x}(i)=1$; any value
  $x(i)\in[1.5\Delta,2.5\Delta)$ becomes $\tilde{x}(i)=2$, etc.

\item[(c)] for negative $x(i)\leq0.5\Delta$ --- any value
  $x(i)\in[-1.5\Delta,0.5\Delta)$ becomes $\tilde{x}(i)=-1$; any value
  $x(i)\in[-2.5\Delta,-1.5\Delta)$ becomes $\tilde{x}(i)=-2$, etc.

\end{itemize}

\begin{figure}
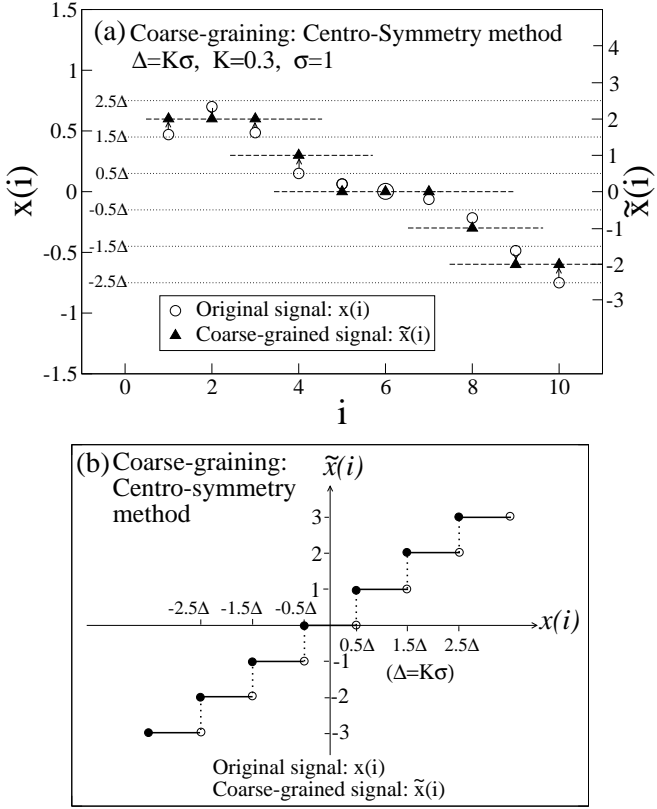

    \includegraphics[width=1\columnwidth]{Fig5a}\vspace*{0.2cm}
    \includegraphics[width=0.8\columnwidth]{Fig5b}
    \caption
    {Schematic illustration of the coarse-graining in the magnitude
      procedure based on the {\it Centro-symmetry method}
      (Eq.~\ref{centro-symmetry-1}).  (a)~Coarse-graining the values
      of an original signal $x(i)$ ($\circ$) with a partition factor
      $\Delta$ to obtain a coarse-grained (symbolic) signal
      $\tilde{x}(i)$ ($\blacktriangle$). $K$ is the partition
      coefficient and $\sigma$ is the standard deviation of
      $x(i)$. Arrows represent the shift from real values $x(i)$ to
      integer values (symbols) $\tilde{x}(i)$ as a result of the
      Centro-symmetry coarse-graining. Horizontal dashed lines
      represent the boundaries of the discretization intervals.
      Horizontal long dashed lines represent the center of each
      discretization interval.  (b)~Illustration of the transition
      from real values of the original signal $x(i)$ to integer values
      (symbols) for the Centro-symmetry coarse-grained signal
      $\tilde{x}(i)$. Solid horizontal lines in (b) represent the
      intervals of $x(i)$ values which become coarse-grained into
      integers (symbols) $\tilde{x}(i)$. Open ($\circ$) and closed
      ($\bullet$) circles indicate opened- and closed-end partition
      intervals for $x(i)$ in the Centro-symmetry transformation,
      e.g., all values of $x(i)\in[0.5\Delta,1.5\Delta)$ are
      transformed into $\tilde{x}(i)=1$, all
      $x(i)\in[-0.5\Delta,0.5\Delta)$ become $\tilde{x}(i)=0$, and all
      values $x(i)\in[-1.5\Delta,-0.5\Delta)$ become
      $\tilde{x}(i)=-1$, etc.}
     \label{fig:illustration of Centro-symmetry Method}

\end{figure}

In Fig.\ref{fig:illustration of Centro-symmetry Method} we
schematically demonstrate the coarse-graining procedure based on the
Centro-symmetry method.

In Fig.\ref{fig:illustration_of Centro-symmetry coarse-grained
signal} we illustrate the effect of the coarse graining in the
magnitude based on the Centro-symmetry method for segments of two
representative signals $x(i)$: one long-range power-law correlated
signal and one power-law  anti-correlated signal. To compare the
effect of the Centro-symmetry method with the effects of the
Symmetry and the Floor methods, we use in
Fig.\ref{fig:illustration_of Centro-symmetry coarse-grained
signal} the same identical signal segments as in
Fig.\ref{fig:illustration_of Floor coarse-grained signal} and
Fig.\ref{fig:illustration of Symmetry coarse-grained signal}.

\begin{figure*}
    \includegraphics[width=1\columnwidth]{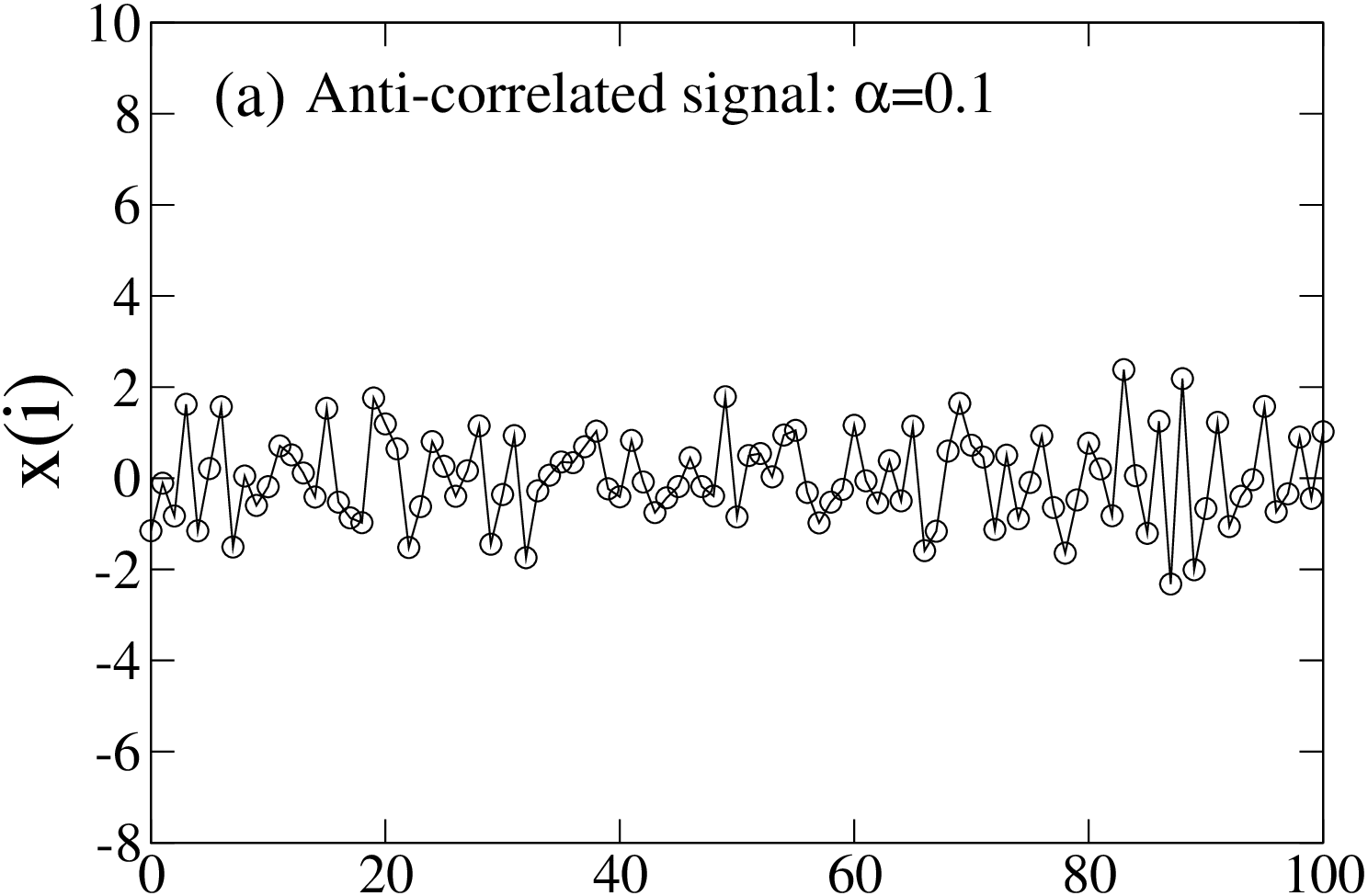}\vspace*{0cm}
    \includegraphics[width=1\columnwidth]{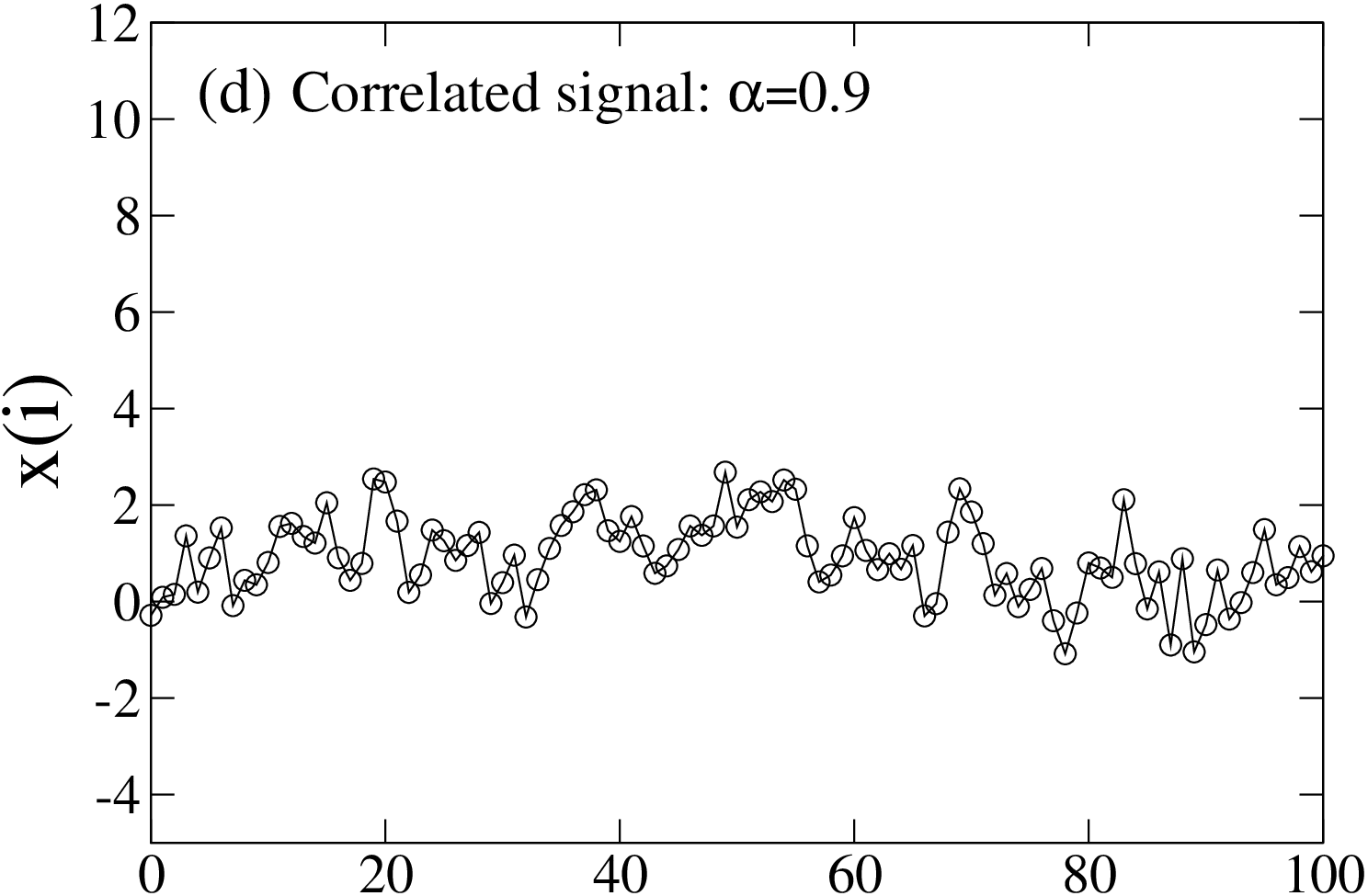}\vspace*{0cm}
    \includegraphics[width=1\columnwidth]{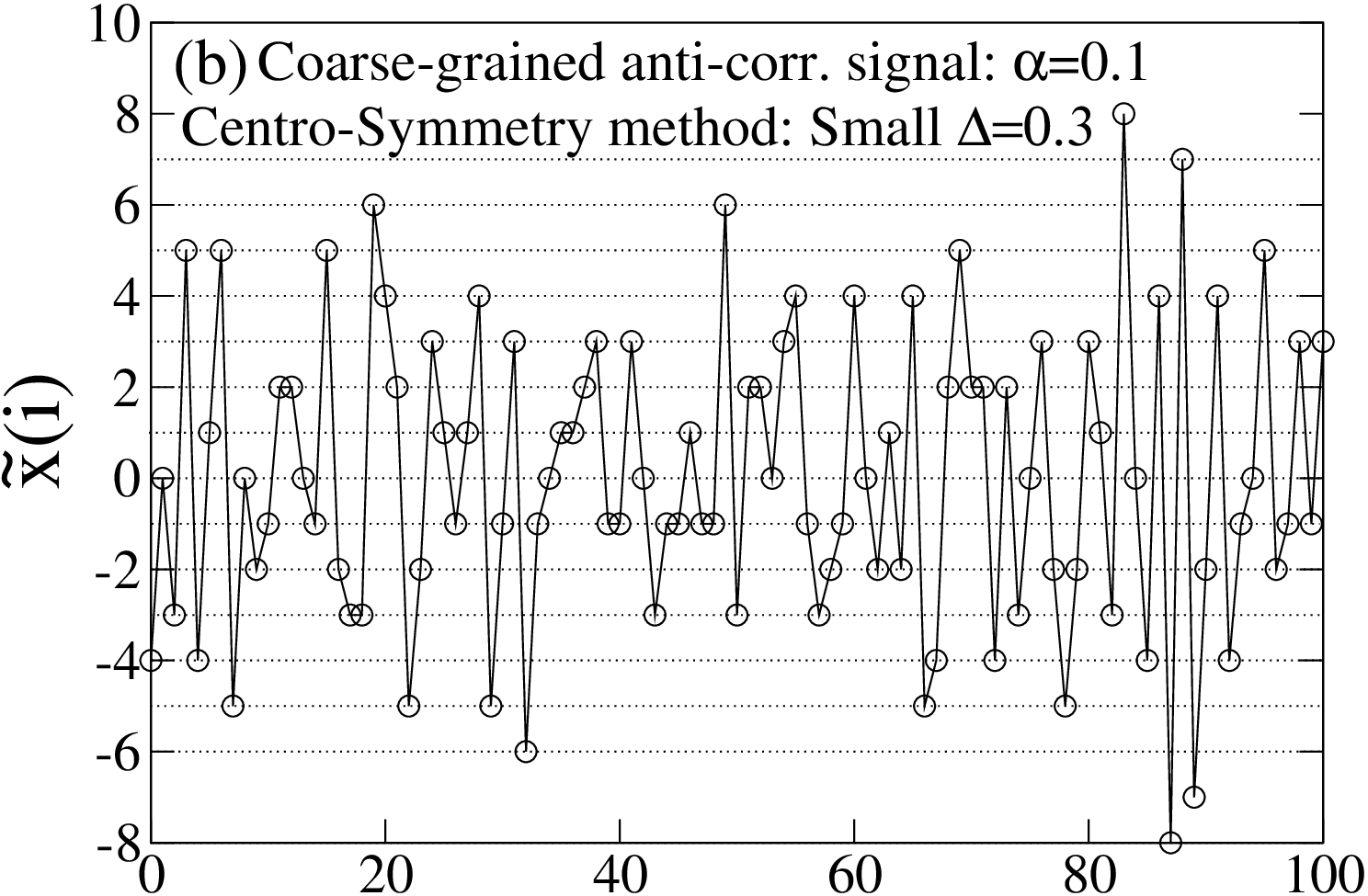}\vspace*{0cm}
    \includegraphics[width=1\columnwidth]{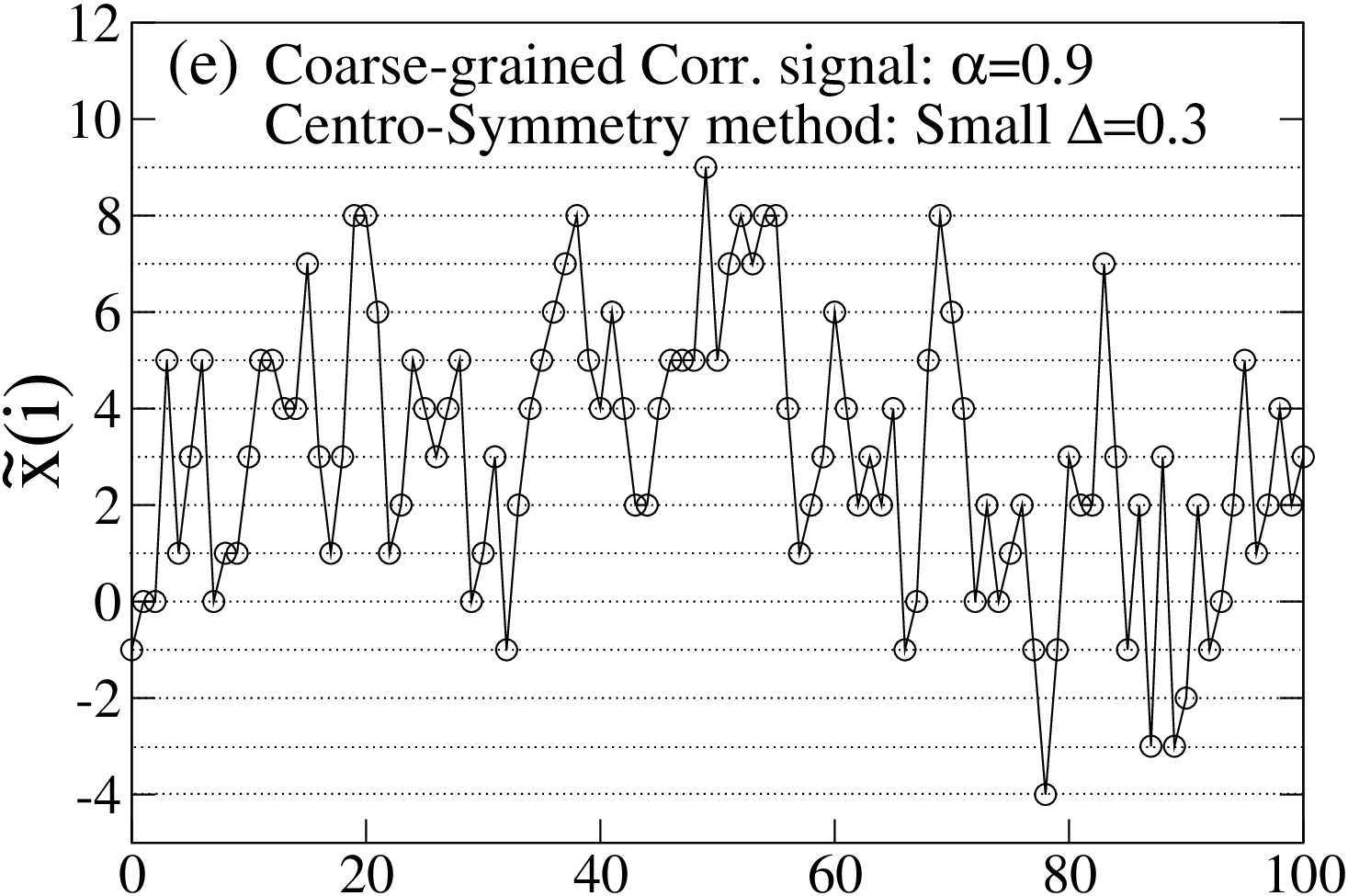}\vspace*{0cm}
    \includegraphics[width=1\columnwidth]{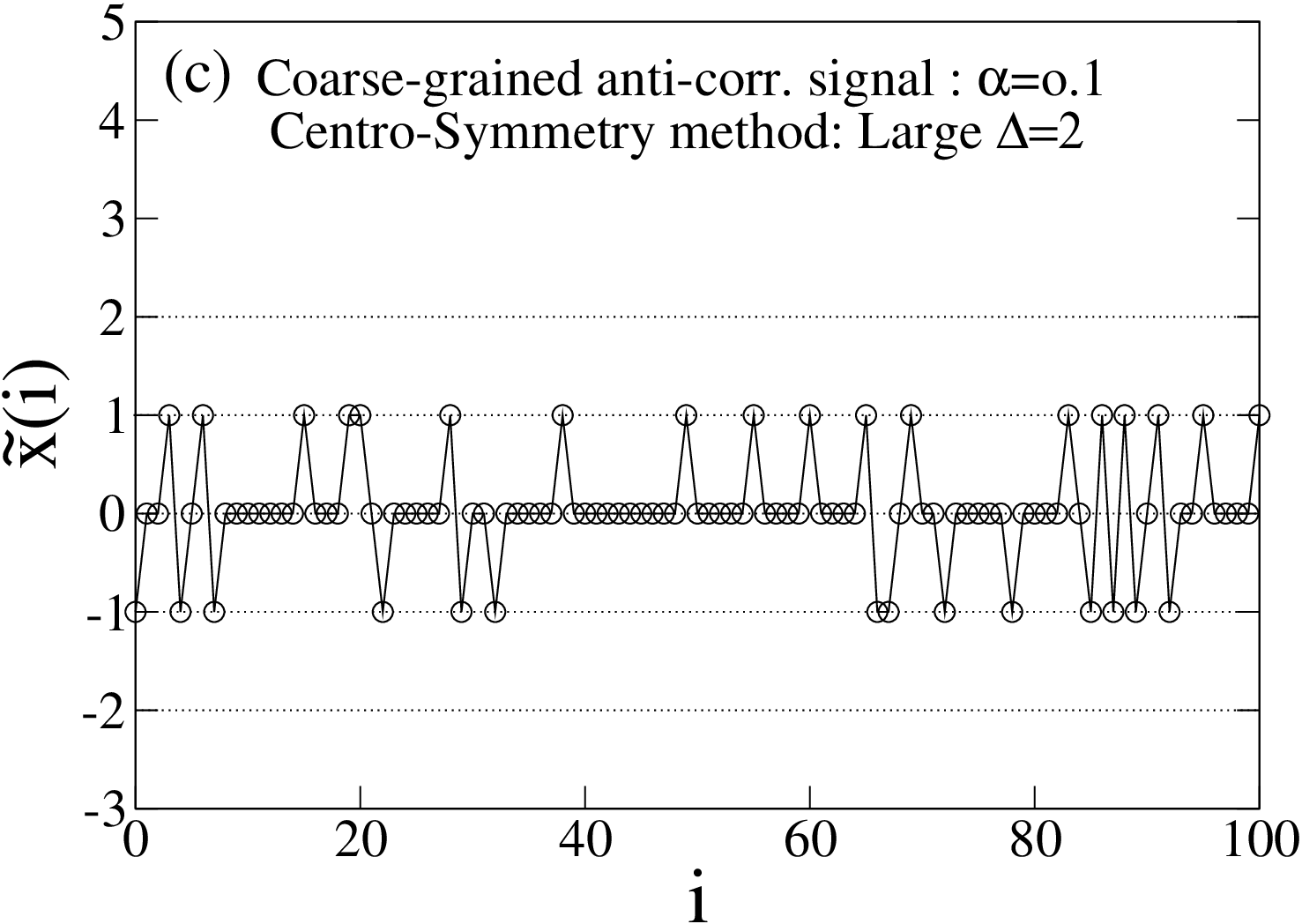}\vspace*{0.4cm}
    \includegraphics[width=1\columnwidth]{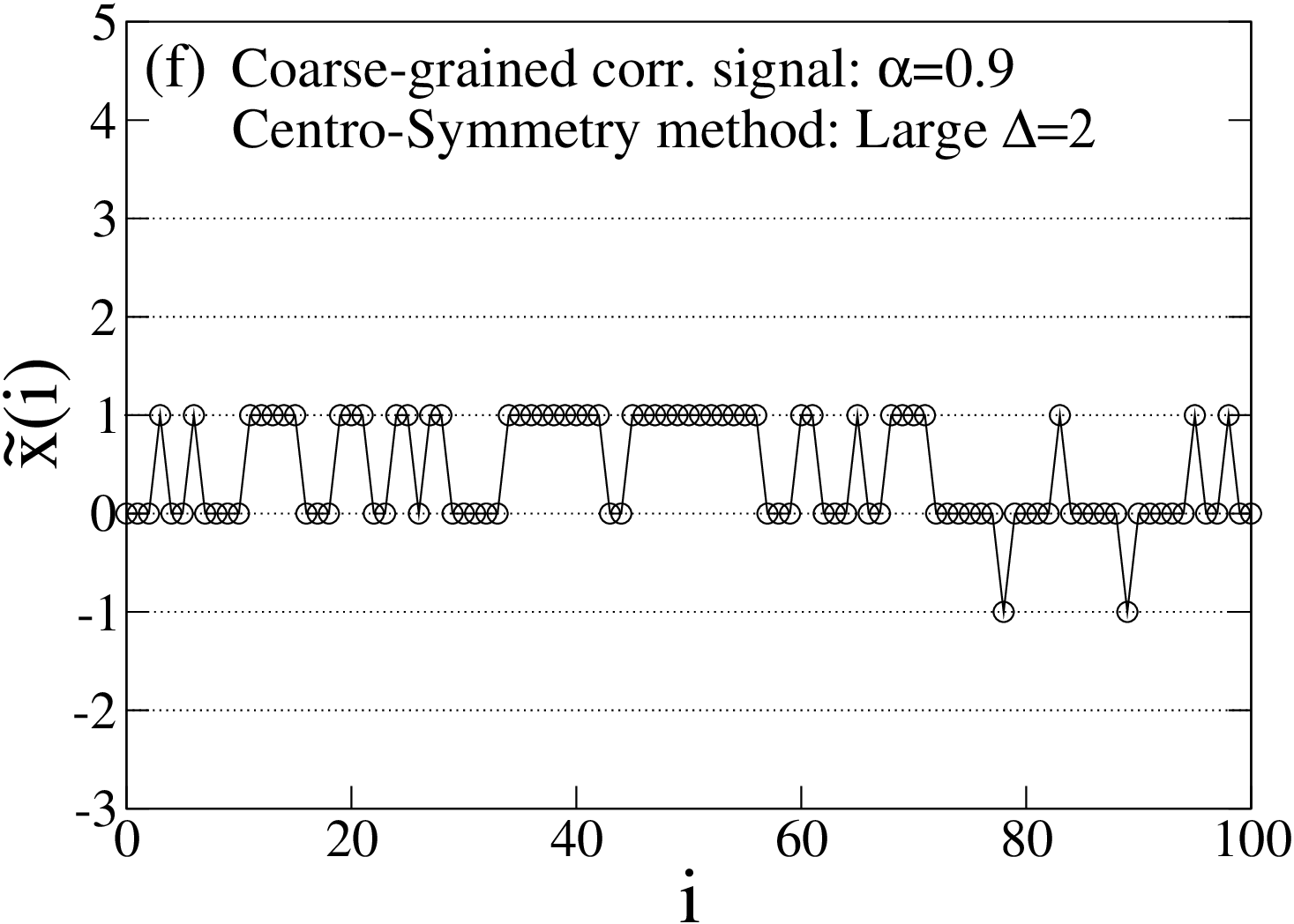}\vspace*{0.4cm}
    \caption { Demonstration of the effect of the {\it Centro-symmetry
        coarse-graining method} on (a)~power-law strongly
      anti-correlated signal $x(i)$ with a DFA scaling exponent
      $\alpha=0.1$ (see Sec. \ref{DFA-method}) and (d)~power-law
      strongly correlated signal $x(i)$ with a DFA scaling exponent
      $\alpha=0.9$.  Both signals in (a) and (d) are identical to
      those shown in Fig.\ref{fig:illustration_of Floor coarse-grained
        signal}a,d, and Fig.\ref{fig:illustration of Symmetry
        coarse-grained signal}a,d, and have standard deviation
      $\sigma=1$.  For small values of the width of the discretization
      partition interval, $\Delta<1$, the Centro-symmetry
      coarse-graining leads to expansion in the magnitude of the
      coarse-grained symbolic signal $\tilde{x}(i)$ along the y-axis
      with much larger standard deviation $\tilde{\sigma}$ compared to
      $\sigma$, shown in (b) and (e).  For very large values of the
      width of the partition interval, $\Delta\gg1$, the
      Centro-symmetry coarse-graining practically leads to binary
      sequences for both correlated and anti-correlated signals, shown
      in (c) and (f). We note that for $\Delta\gg1$ the standard
      deviation $\tilde{\sigma}$ of $\tilde{x}(i)$ is significantly
      smaller than $\sigma$ of $x(i)$ (see Fig.\ref{fig:Effect of
        Centro-Symmetric method coarse-graining on DFA}a,c).  We also
      note, that for both $\Delta<1$ and $\Delta>1$ the
      Centro-symmetry coarse-graining leads to a larger fraction of
      values $\tilde{x}(i)=0$ compared to the Symmetry and Floor
      coarse-graining methods.  }
     \label{fig:illustration_of Centro-symmetry coarse-grained signal}
\end{figure*}

Note, that for values of the width of the partition intervals
$\Delta<1$, the Centro-symmetry coarse-grained signals $\tilde{x}(i)$
expand along the y-axis (Fig.\ref{fig:illustration_of Centro-symmetry
  coarse-grained signal}b,e), an effect similar to the Floor and
Symmetry methods (Fig.\ref{fig:illustration_of Floor coarse-grained
  signal}b,e and Fig.\ref{fig:illustration of Symmetry coarse-grained
  signal}b,e).

However, in contrast to the Symmetry method, for $\Delta>1$ the
Centro-symmetry coarse-graining leads to a relatively larger
proportion of zero values for the symbolic signal $\tilde{x}(i)$
(Fig.\ref{fig:illustration_of Centro-symmetry coarse-grained
  signal}c,f).

\subsubsection{Coarse-graining in time}
\label{Coarse-graining time method}
In addition to coarse-graining the magnitude of a signal (as discussed
in the subsection above), we also consider coarse-graining in time.

Under the coarse-graining in time procedure, we divide the signal
$x(i)$ into segments corresponding to consecutive non-overlapping time
intervals of size $\Delta$, and we replace all data points in each
segment by a single data point with a value equal to the average of
the signal in this segment.  Given a signal $x(i) (i=1,...,N_{\rm
  max})$, we construct a coarse-grained in time series $\tilde{x}(j)
(j=1,...,\lfloor N_{\rm max}/\Delta \rfloor)$, using the following
rule:

\begin{equation}
  \tilde{x}(j)\equiv\frac{1}{\Delta}\sum\limits_{i=(j-1)\Delta+1}^{j\Delta} x(i),~~1\leq j\leq
  \left \lfloor\frac{N_{\rm max}}{\Delta}\right
  \rfloor,
  \label{coarse-graining in time}
\end{equation}

where the floor function $\left\lfloor N_{\rm max}/\Delta \right
\rfloor$ indicates the largest integer number less or equal to $N_{\rm
  max}/\Delta$. Thus, the index $j$ of the coarse-grained in time
signal $\tilde{x}(j)$ denotes the index of the consecutive
non-overlapping time intervals of size $\Delta$ which partition the
original signal $x(i)$. The length of signal $\tilde{x}(j)$ is
determined by the maximum number of time intervals of size $\Delta$
which can fit in the length $N_{\rm max}$ of the original signal
$x(i)$. Thus $\Delta$ plays the role of a scale factor in
`renormalizing' (coarse-graining) the original signal.

\begin{figure}
    \includegraphics[width=1\columnwidth]{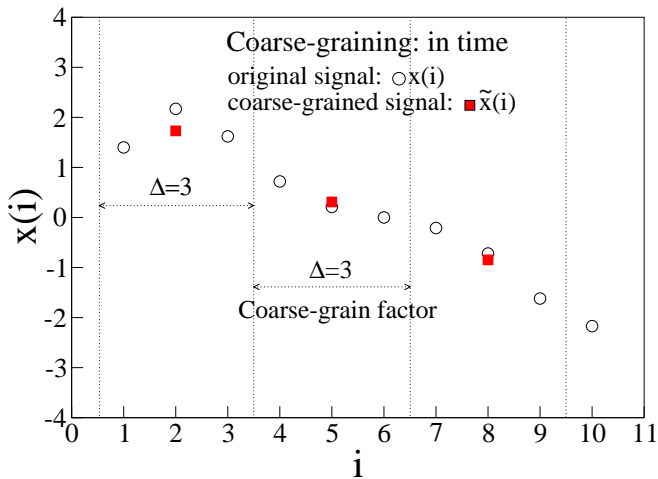}\vspace*{0.2cm}
    \caption
    {Schematic illustration of the {\it coarse-graining in time}
      procedure. Coarse graining the values of an original signal
      $x(i)$ ($\circ$) with a scale factor (time interval) $\Delta=3$
      to obtain the coarse-grained signal $\tilde{x}(j)$ (full
      squares). In each time interval $\Delta$ the all values of
      $x(i)$ are replaced by their average $\tilde{x}(j)$
      (\ref{coarse-graining in time}).  }
     \label{fig:Coarse-graining in time}
\end{figure}

In Fig.\ref{fig:Coarse-graining in time} we schematically demonstrate
the coarse-graining in time method. For scale factor $\Delta=1$ the
coarse-graining in time rule in Eq.{\ref{coarse-graining in time}}
reproduces the original signal $\tilde{x}(j)\equiv x(i)$ where j=i,
i.e., coarse-graining with an unit scale factor does not coarse-grain
the signal $x(i)$.

\begin{figure}
    \includegraphics[width=1\columnwidth]{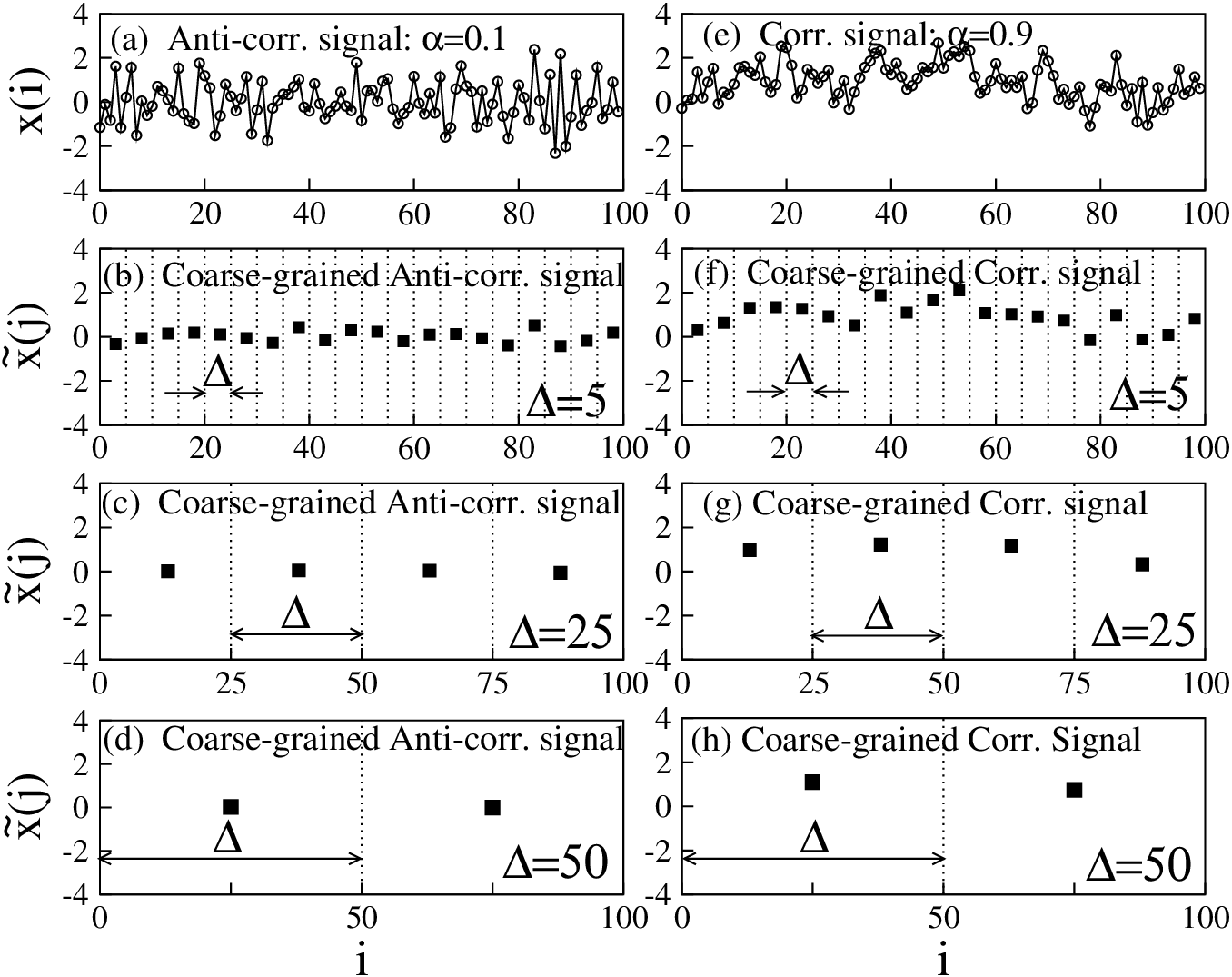}\vspace*{0.4cm}

    \caption {Demonstration of the effect of the {\it coarse-graining
        in time} method on (a) power-law strongly anti-correlated
      signal $x(i)$ with a DFA scaling exponent $\alpha=0.1$ (see
      Sec. \ref{DFA-method}) and (e) power-law strongly correlated
      signal $x(i)$ with a DFA scaling exponent $\alpha=0.9$. Both
      signals in (a) and (e) are identical to those shown in
      Fig.\ref{fig:illustration_of Floor coarse-grained signal}a,d,
      Fig.\ref{fig:illustration of Symmetry coarse-grained signal}a,d,
      and Fig.\ref{fig:illustration_of Centro-symmetry coarse-grained
        signal}a,d, and have standard deviation $\sigma=1$.  With
      increasing the scale factor $\Delta$, the variability and the
      standard deviation is reduced for both anti-correlated (show in
      (b)-(d)) and for correlated signals (shown in (f)-(h)). For the
      same scale factor $\Delta$, this effect is stronger for
      anti-correlated signals (compare (b) to (f)).  }
     \label{fig:Example of coarse-graining in time}
\end{figure}

In Fig.\ref{fig:Example of coarse-graining in time} we demonstrate
the effect of the coarse-graining in time method for two
representative signals $x(i)$: one with long-range power-law
correlations and one with power-law anti-correlations. For
comparison of the different coarse-graining methods, the signals
we show in Fig.\ref{fig:Example of coarse-graining in time} are
identical to the signals shown in 
Fig.\ref{fig:illustration_of Floor coarse-grained signal},
Fig.\ref{fig:illustration of Symmetry coarse-grained signal} and 
Fig.\ref{fig:illustration_of Centro-symmetry coarse-grained signal}.
 With increasing the scale factor $\Delta$ the variability
 (standard deviation $\sigma$) of
the coarse-grained in time signal $\tilde{x}(j)$ monotonically
decreases (Fig.\ref{fig:Example of coarse-graining in time}). For
the same value of the scale factor $\Delta$, the reduction of 
variability is more pronounced for anti-correlated signals 
as compared to positively correlated signals.

\subsection{DFA method}
\label{DFA-method}
Using the Fourier filtering method presented in
the section \ref{Fourier filtering method}, we generate stationary
signals with long-range power-law correlations. To quantify these
correlations and how they change after different coarse-graining
procedures, we apply the DFA method \cite{CKDFA1}.

We briefly introduce the DFA method, which involves the following
steps \cite{CKDFA1}:

(i) A given signal $u(i)$ ($i=1,..,N$, where $N$ is the length of the
signal) is integrated to obtain the random walk profile
$y(k)\equiv\sum_{i=1}^{k}[u(i)-\langle u \rangle]$, where $\langle u
\rangle$ is the mean of $u(i)$.
 
(ii) The integrated signal $y(k)$ is divided into boxes of equal
length $n$.
 
(iii) In each box of length $n$ we fit $y(k)$ using a polynomial
function of order $\ell$ which represents the {\it trend\/} in that
box. The $y$ coordinate of the fit curve in each box is denoted by
$y_n(k)$. When a polynomial fit of order $\ell$ is used, we denote the
algorithm as DFA-$\ell$. Note that, due to the integration procedure
in step (i), DFA-$\ell$ removes polynomial trends of order $\ell-1$ in
the original signal $u(i)$.

(iv) The integrated profile $y(k)$ is detrended by subtracting the
local trend $y_n(k)$ in each box of length $n$:
\begin{equation}
 Y(k)\equiv~y(k)-y_n(k).
\label{F2}
\end{equation} 
 
(v) For a given box length $n$, the root-mean-square (rms) fluctuation
function for this integrated and detrended signal is calculated:
\begin{equation}
 F(n)\equiv\sqrt{{1\over {N}}\sum_{k=1}^{N}[Y(k)]^2}.
\label{F2}
\end{equation} 

(vi) The above computation is repeated for a broad range of box
lengths $n$ (where $n$ represents a specific space or time scale) to
provide a relationship between $F(n)$ and $n$.

\begin{figure}
\vspace*{0.truein} \centerline{
  \includegraphics[width=0.85\columnwidth]{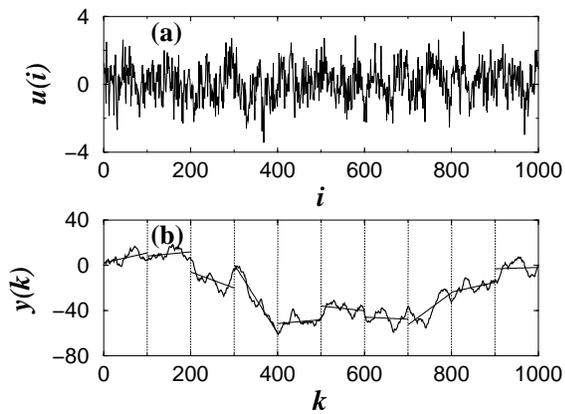}}
\caption{(a) The correlated signal $x(i)$. (b) The integrated signal:
  $y(k)=\sum_{i=1}^k[x(i)-\langle x \rangle]$. The vertical dotted
  lines indicate a box of size $n=100$, the solid straight lines
  segments are the estimated linear ``trend'' in each box by
  least-squares fit.}
\label{f.dfa-box}
\end{figure}

A power-law relation between the average root-mean-square fluctuation
function $F(n)$ and the box size $n$ indicates the presence of
scaling: $F(n) \sim n^{\alpha}$. The fluctuations can be characterized
by a scaling exponent $\alpha$, a self-similarity parameter which
represents the long-range power-law correlation properties of the
signal. If $\alpha=0.5$, there is no correlation and the signal is
uncorrelated (white noise); if $\alpha < 0.5$, the signal is
anti-correlated; if $\alpha >0.5$, the signal is correlated. For
stationary signals, the scaling exponent $\alpha$ relates linearly to
the scaling exponent $\beta$ of the power spectrum $S(q)$
(Eq.\ref{M1}) and to the correlation exponent $\gamma$ of the
auto-correlation function $C(n)$ (Eq.\ref{M3}) as:
\begin{equation}
\alpha=\frac{\beta+1}{2}=\frac{2-\gamma}{2},
\label{M4}
\end{equation}
where $0<\gamma<1$.

We note that for anti-correlated signals, the scaling exponent
obtained from the DFA method overestimates the true correlations at
small scales \cite{kun}. To avoid this problem, one needs first to
integrate the original anti-correlated signal and then apply the DFA
method \cite{kun}. The correct scaling exponent can thus be obtained
from the relation between $n$ and $F(n)/n$ (instead of $F(n)$).
 In the following sections, we first integrate the signals under
 consideration, then apply DFA-2 to remove linear trends in these
integrated signals. In order to provide a more accurate estimate of
$F(n)$, the largest box size $n$ we use is $N_{\rm max}/10$, where
$N_{\rm max}$ is the total number of points in the signal.

We compare the results of the DFA method obtained from the
coarse-grained signals $\tilde{x}(i)$ with those obtained from the
stationary signal $x(i)$, and examine how the scaling properties of
a detrended fluctuation function $F(n)$ change when introducing
different types of coarse-graining.

\section{Results}\label{Results}
\subsection{Coarse-graining in magnitude}
\label{Result-magnitude}

(i) {\it Floor Method}. We first consider how symbolic coarse-graining
in the magnitude based on the Floor method (see Section
\ref{Coarse-graining magnitude methods}, Eq.\ref{floor-1} and
Fig.\ref{fig:illustration_of Floor Method}) affects the scaling
behavior of long-range power-law correlated signals. We systematically
investigate the effect of this type of coarse-graining on both
anti-correlated and correlated signals.

\begin{figure*}
    \includegraphics[width=0.85\columnwidth]{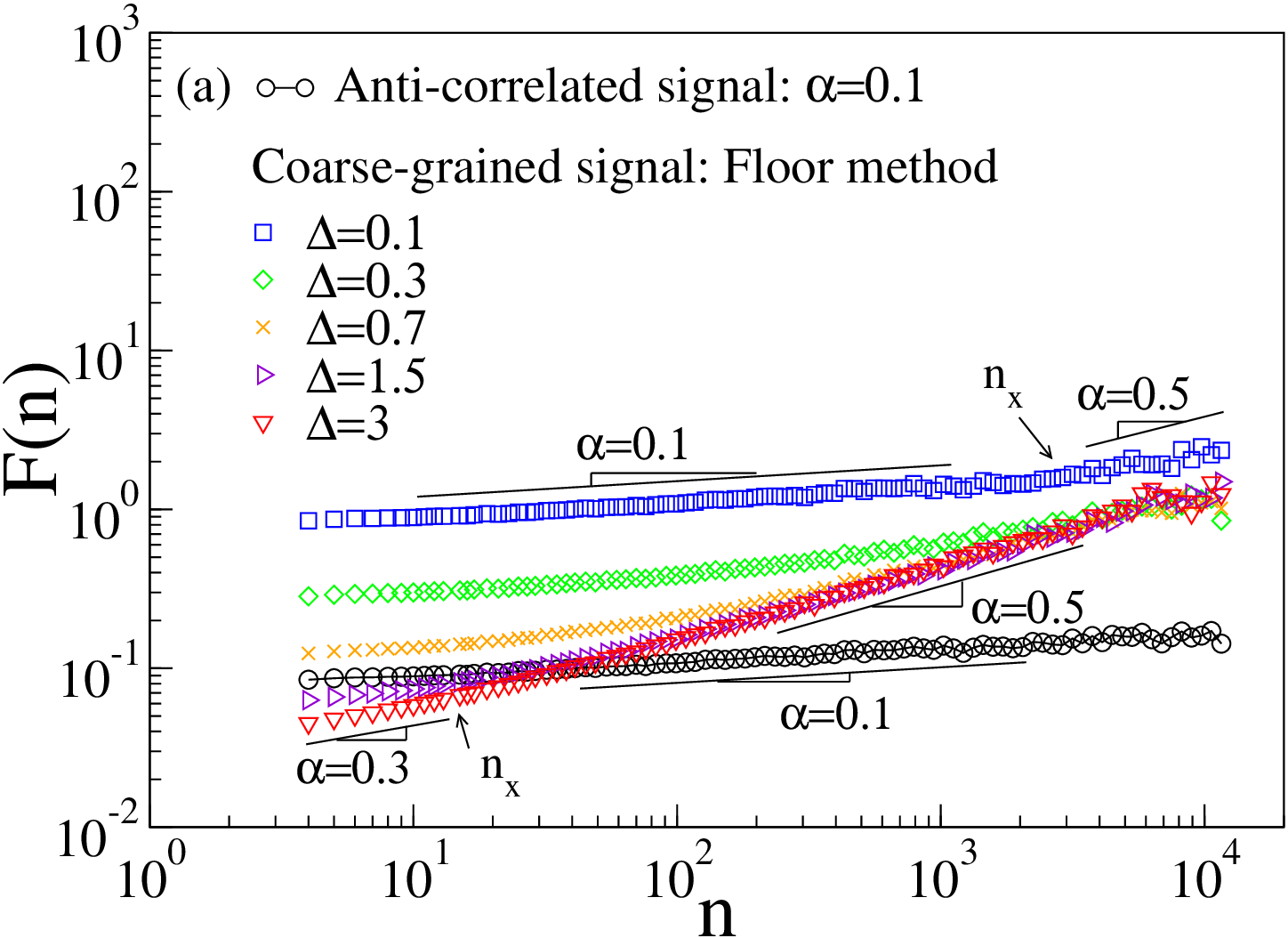}\vspace*{0cm}
    \includegraphics[width=0.85\columnwidth]{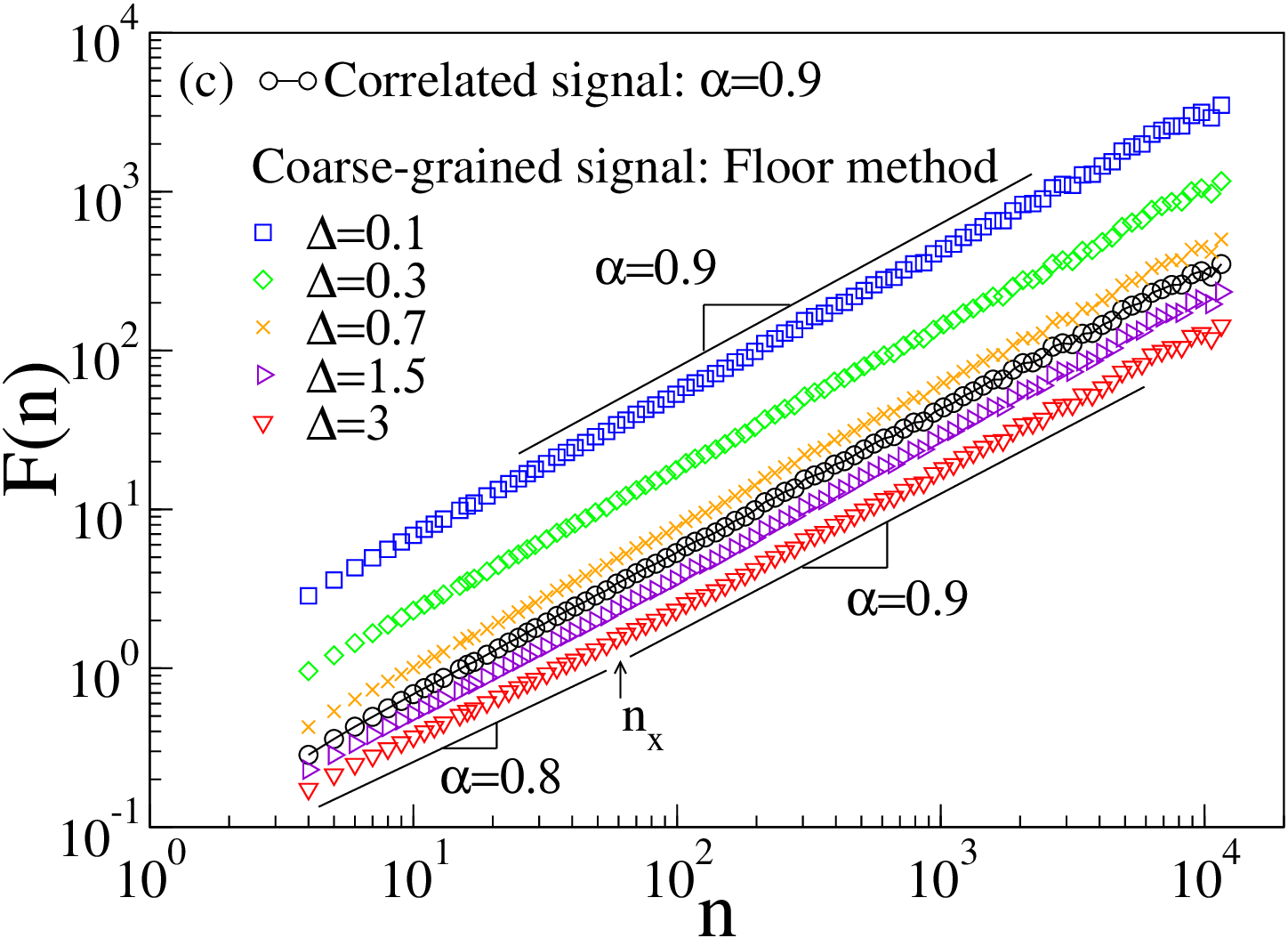}\vspace*{0cm}
    \includegraphics[width=0.85\columnwidth]{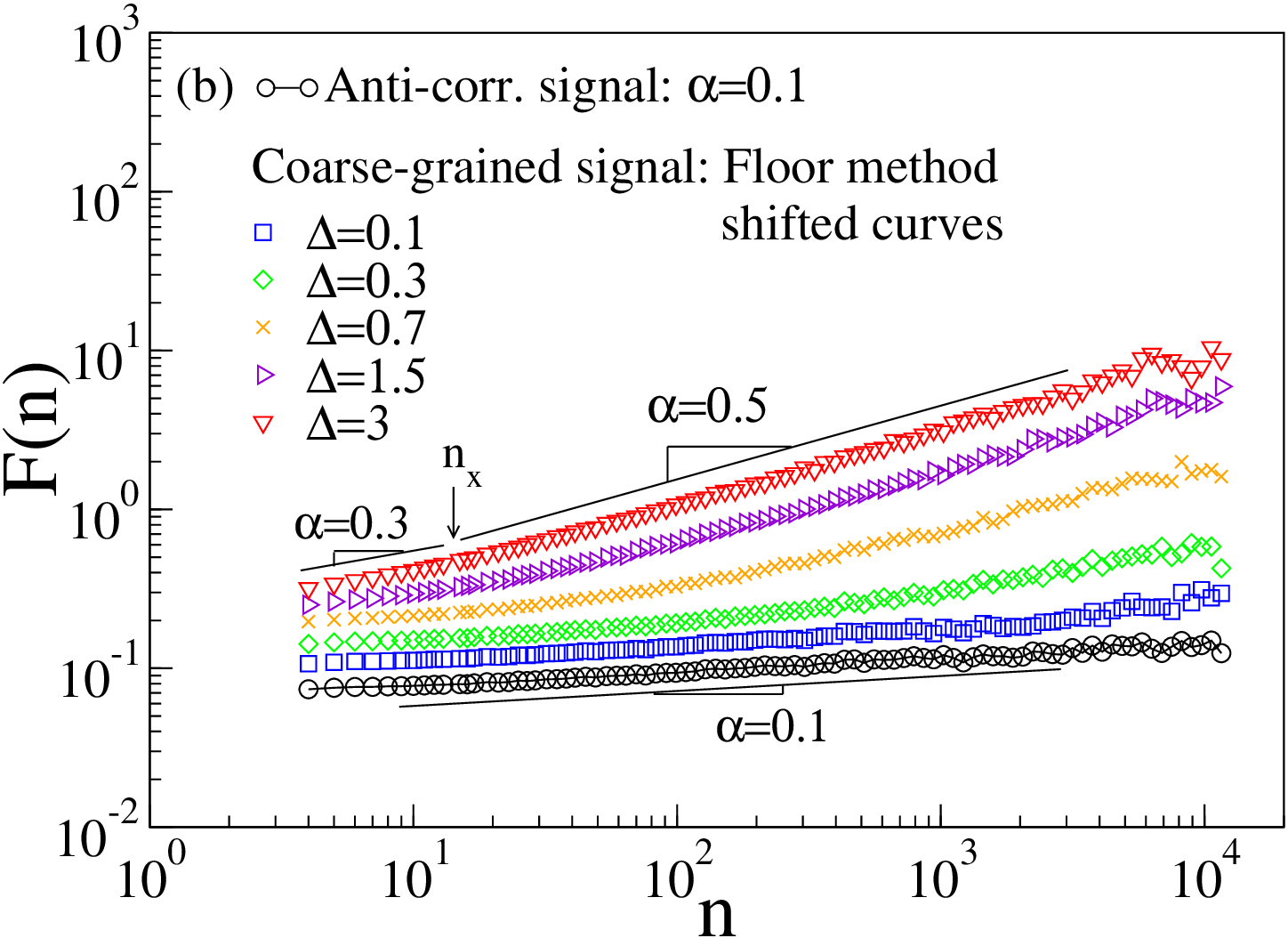}\vspace*{0cm}
    \includegraphics[width=0.85\columnwidth]{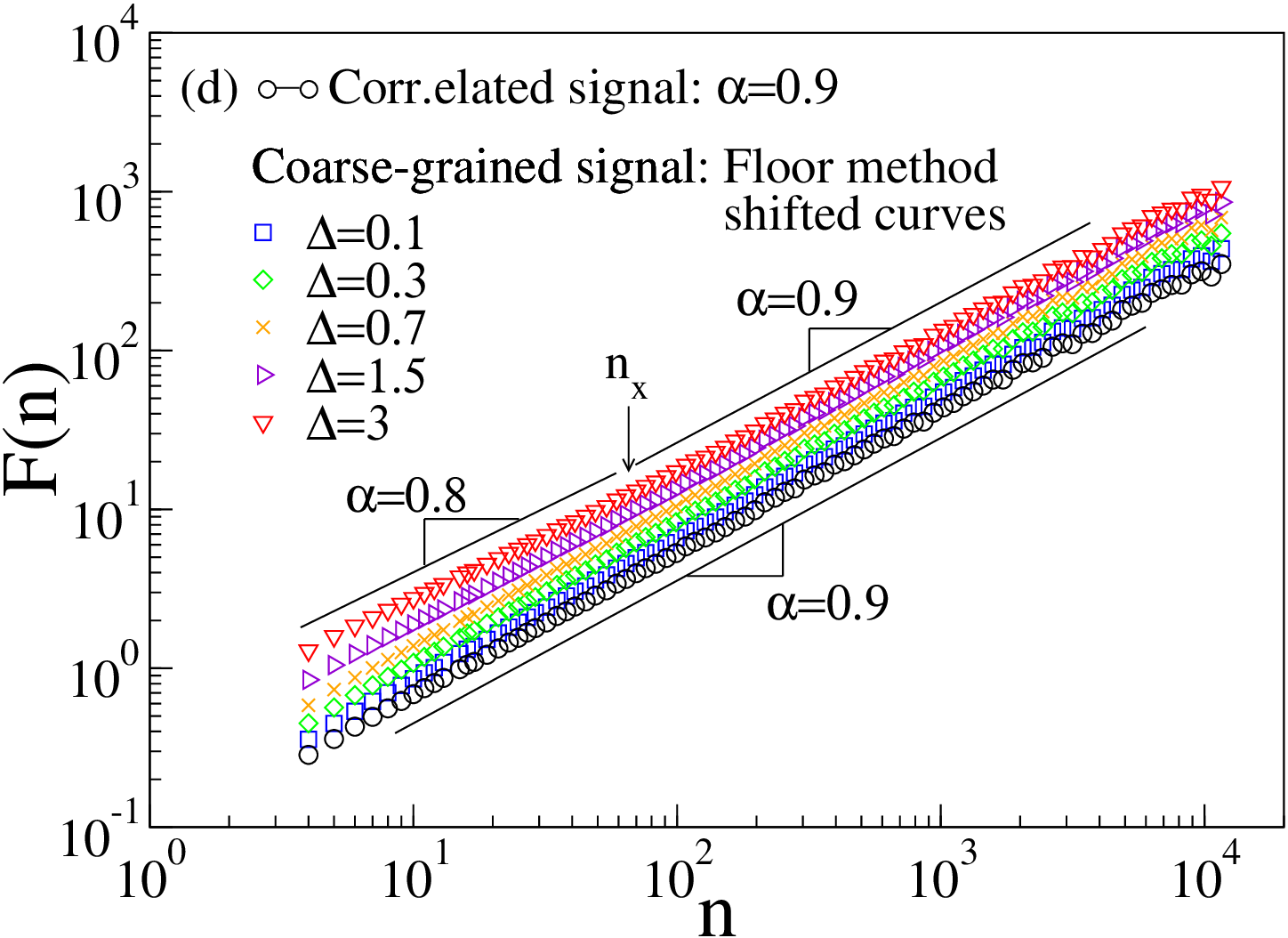}\vspace*{0cm}
    \includegraphics[width=0.85\columnwidth]{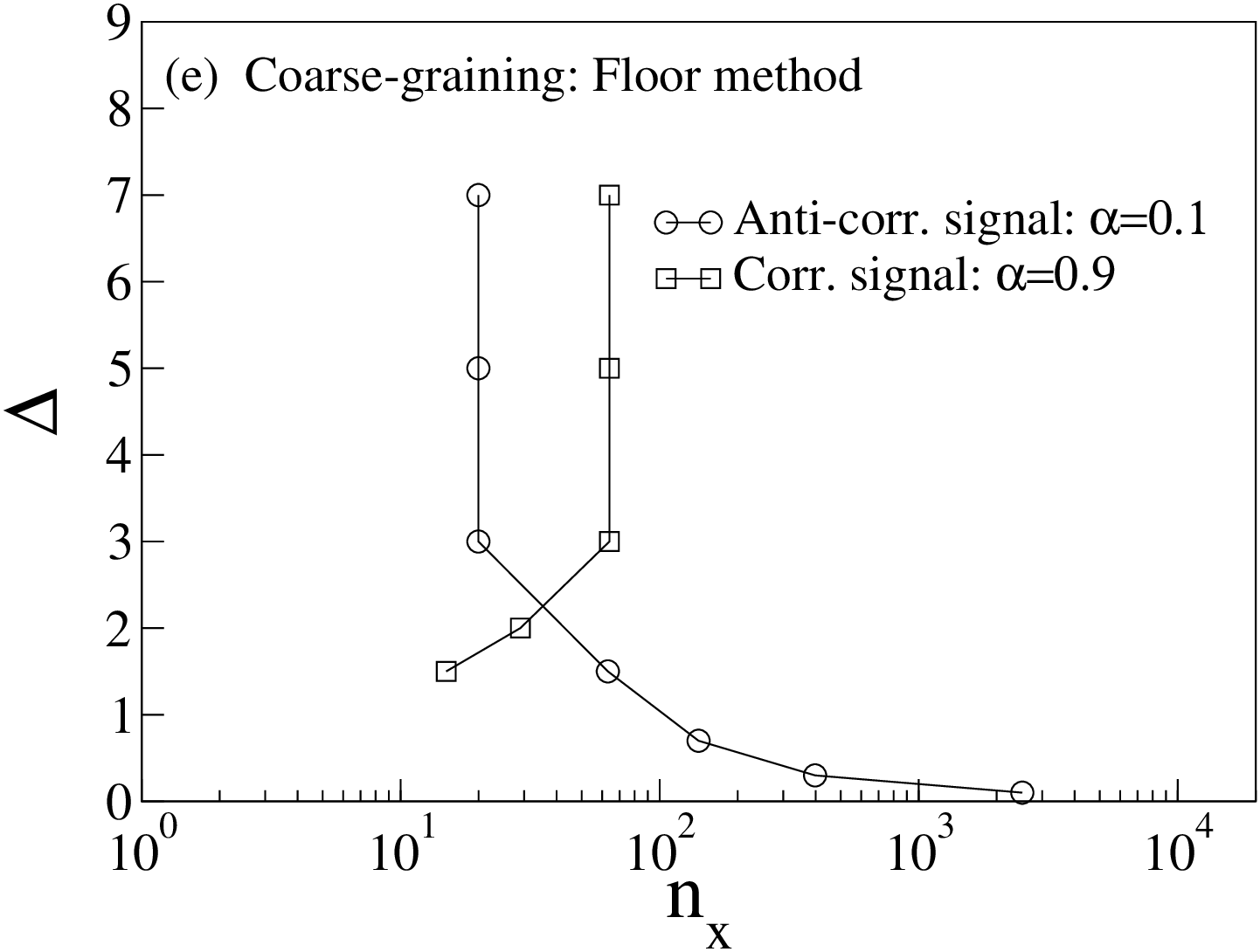}
    \caption { Effect of the {\it Floor coarse-graining method} on the
      scaling behavior of long-range power-law correlated signals.
      Scaling curves $F(n)$ obtained using the DFA-2 method of (a) an
      anti-correlated signal with scaling exponent $\alpha=0.1$, and
      (c) a positively correlated signal with $\alpha=0.9$ before and
      after coarse-graining with different values of the partition
      parameter $\Delta$. The same signals as shown in
      Fig.\ref{fig:illustration_of Floor coarse-grained signal}a, d
      are used for the analysis. For $\Delta<1$, the scaling curves
      $F(n)$ of the coarse-grained signals $\tilde{x}(i)$ shift up
      compared to $F(n)$ of the original signal $x(i)$ due to the
      expansion of the interface of $\tilde{x}(i)$ compared to $x(i)$
      (shown in Fig.\ref{fig:illustration_of Floor coarse-grained
        signal}b, e). With increasing $\Delta>1$, the interface of
      $\tilde{x}(i)$ shrinks and $F(n)$ shifts down. For the
      anti-correlated signal in (a), the Floor coarse-graining leads
      to a crossover from anti-correlated ($\alpha=0.1$) to random
      behavior ($\alpha=0.5$) at a crossover scale $n_x$. For
      $\Delta<1$ the crossover $n_x$ corresponds to large scales, and
      moves to intermediate and small scales with increasing
      $\Delta$. In contrast to anti-correlated signals, the Floor
      coarse-graining with small $\Delta<1$ has no effect on the
      scaling of positively correlated signals, while for large
      $\Delta>1$ there is a crossover at small scales to weaker
      correlations ($\alpha=0.8$). (b)~and~(d) show the same results
      as in (a) and (c) with scaling curves $F(n)$ vertically shifted
      to visualize the transition of the crossover $n_x$ for different
      values of $\Delta$. We determine the crossover scale $n_x$ based
      on the difference $\epsilon(n)= {\rm log}_{10}F(n)-{\rm
        log}_{10}F_{\rm fit}(n)$ between the scaling curve $F(n)$ of
      the coarse-grained signal and the linear fitting line $F_{\rm
        fit}(n)$ for the range of scales not affected by the
      coarse-graining: we chose $n_x$ as the scale where
      $\epsilon(n)=0.04$.  Dependence of the crossover scale $n_x$ on
      the coarse-graining parameter $\Delta$ for the anti-correlated
      and positively correlated signals shown in (a) and (c). Note,
      that for very rough Floor coarse-graining with $\Delta \ge 3$,
      the crossover scale $n_x$ does not change for both
      anti-correlated and positively correlated signals.}

    \label{fig:Effect of Floor method coarse-graining on DFA}
\end{figure*}

{\it Anti-correlated signals:} We find that the scaling behavior of
anti-correlated signals is strongly affected by the Floor
coarse-graining.  In Fig.\ref{fig:Effect of Floor method
  coarse-graining on DFA}(a,b,e) we present the results of Floor
coarse-graining for a strongly anti-correlated signal characterized by
a DFA scaling exponent $\alpha=0.1$. For small values of the width of
the partition interval $\Delta<1$ the interface of the coarse-grained
signal $\tilde{x}(i)$ expands along the $y$-axis, as demonstrated in
Fig.\ref{fig:illustration_of Floor coarse-grained signal}b, leading to
an increase and a vertical shift up of the DFA fluctuation function
$F(n)$ (Fig.\ref{fig:Effect of Floor method coarse-graining on DFA}a).
Further, with increasing values of $\Delta>1$ the interface of the
coarse-grained anti-correlated signal shrinks, i.e., the standard
deviation decreases (as shown in Fig.\ref{fig:illustration_of Floor
  coarse-grained signal}c), and the scaling curve $F(n)$ shifts
vertically down (Fig.\ref{fig:Effect of Floor method coarse-graining
  on DFA}a).

In addition, the Floor coarse-graining leads to a crossover from
anti-correlated ($\alpha=0.1$) to random behavior ($\alpha=0.5$) in
the scaling curve $F(n)$ at large scales $n$. This crossover is
indicated by $n_x$ in Fig.\ref{fig:Effect of Floor method
  coarse-graining on DFA}a.  With increasing values of $\Delta$ the
crossover scale $n_x$ moves from large to intermediate and small
scales $n$.  This crossover transition from large to small scales $n$
is shown in Fig.\ref{fig:Effect of Floor method coarse-graining on
  DFA}b where for clarity the scaling curves $F(n)$ are vertically
shifted up (relative to the original signal $x(i)$ with $\alpha=0.1$)
in the order of increasing values of $\Delta$. We note, that for very
large values of $\Delta$, the anti-correlated behavior of the original
signal is not preserved even at small scales $n$. This crossover
behavior in the scaling of anti-correlated signals reflects the
different effect of the Floor coarse-graining on the temporal
organization of the original signal values at large and small scales.

For fine coarse-graining with $\Delta<<1$ the relative disposition of
data values in the coarse-grained signal $\tilde{x}(i)$ is preserved
compared to the original anti-correlated signal $x(i)$, as shown in
Fig.\ref{fig:illustration_of Floor coarse-grained signal}a and
Fig.\ref{fig:illustration_of Floor coarse-grained signal}b. As a
result, the scaling exponent $\alpha$ characterizing the curve $F(n)$
for $\tilde{x}(i)$ is almost the same as for the original
anti-correlated signal $x(i)$ --- for example, $F(n)$ for $\Delta=0.1$
is practically parallel to $F(n)$ for the original signal except for
slight deviation only at very large scales $n$ (see
Fig.\ref{fig:Effect of Floor method coarse-graining on DFA}a,b). For
anti-correlated signals, increasing the width of the partition
interval $\Delta$ leads to stronger coarse-graining effect at larger
scales while at small time scales, the relative disposition of
neighboring coarse-grained data values $\tilde{x}(i)$ is preserved
similar to the disposition of the corresponding original values
$x(i)$, so that at small and intermediate scales the anti-correlations
are preserved.  Thus, the Floor coarse-graining method leads to a
systematic change in the correlation properties of the original signal
$x(i)$, with a crossover from anti-correlated behavior (characterized
by exponent $\alpha<0.5$) to random behavior (with $\alpha=0.5$), and
with a crossover scale $n_x$ shifting from large to intermediate and
small scales $n$ as $\Delta$ increases.

For very rough coarse-graining, with partition parameter
$\Delta\geqslant 3$, we find that even at small scales the original
anti-correlated behavior is not preserved and the exponent
$\alpha=0.1$ can not be recovered (see Fig.\ref{fig:Effect of Floor
  method coarse-graining on DFA}b). We note that because the original
signal $x(i)$ has a standard deviation $\sigma=1$, using a partition
parameter $\Delta\geqslant3$ leads practically to a binary signal
$\tilde{x}(i)$ (see Fig.\ref{fig:illustration_of Floor coarse-grained
  signal}c). Thus, for $\Delta>3$ the scaling curve $F(n)$ does not
change any more and remains the same as for $\Delta=3$ (the crossover
scale $n_x$ also does not change for $\Delta\geqslant3$, see
Fig.\ref{fig:Effect of Floor method coarse-graining on DFA}(e)).

The change we observe in the scaling behavior of power-law
anti-correlated signals $x(i)$ after Floor coarse-graining is similar
to the effect of adding noise to each data point in $x(i)$ or of
adding a percentage of random spikes of a given amplitude to the
original signal. It was previously reported that random spikes alter
the scaling curve $F(n)$ of an anti-correlated signal leading to a
crossover from anti-correlated ($\alpha<0.5$) behavior to random
behavior ($\alpha=0.5$) \cite{Zhichen_PRE_2002}.  It was also reported
that with increasing the amplitude of the spikes and their
concentration in the original signal the crossover scale $n_x$
gradually moves from large to intermediate and small scales (see
Fig.3d in \cite{Zhichen_PRE_2002}), an effect identical to our results
shown in Fig.\ref{fig:Effect of Floor method coarse-graining on
  DFA}a,b. Indeed, with increasing the partition parameter $\Delta$
the difference between the values $x(i)$ in the original
anti-correlated signal and the corresponding integer (symbolic) values
$\tilde{x}(i)$ in the coarse-grained signal also increases,
effectively equivalent to adding random noise. The rougher the
coarse-graining (i.e., larger $\Delta$) the higher the percentage and
amplitude of the effectively added noise, as more data points in the
original signal $x(i)$ are affected by the coarse-graining procedure
(higher percentage), while at the same time the change in the relative
(compared to the standard deviation of the signal) increments in
$x(i)$ values is larger with increasing $\Delta$. Thus, the effect of
Floor coarse-graining on the scaling of anti-correlated signals is
equivalent to a superposition of the scaling curve $F(n)$ of the
original signal with the scaling curve $F_{\eta}(n)$ of a white noise
signal ($\alpha=0.5$). Increasing the amplitude and standard deviation
of the added white noise signal leads to an increase (a vertical shift
up) in $F_{\eta}(n)$, and thus, a shift of the intersection (i.e., the
crossover scale $n_x$) between the $F(n)$ curves of the white noise
and of the anti-correlated signal from large to smaller scales
$n$. The role of such a superposition rule on the crossover behavior
of correlated signals with nonstationarities has been discussed in
detail in \cite{kun,Zhichen_PRE_2002}.

\begin{figure*}
    \includegraphics[width=0.85\columnwidth]{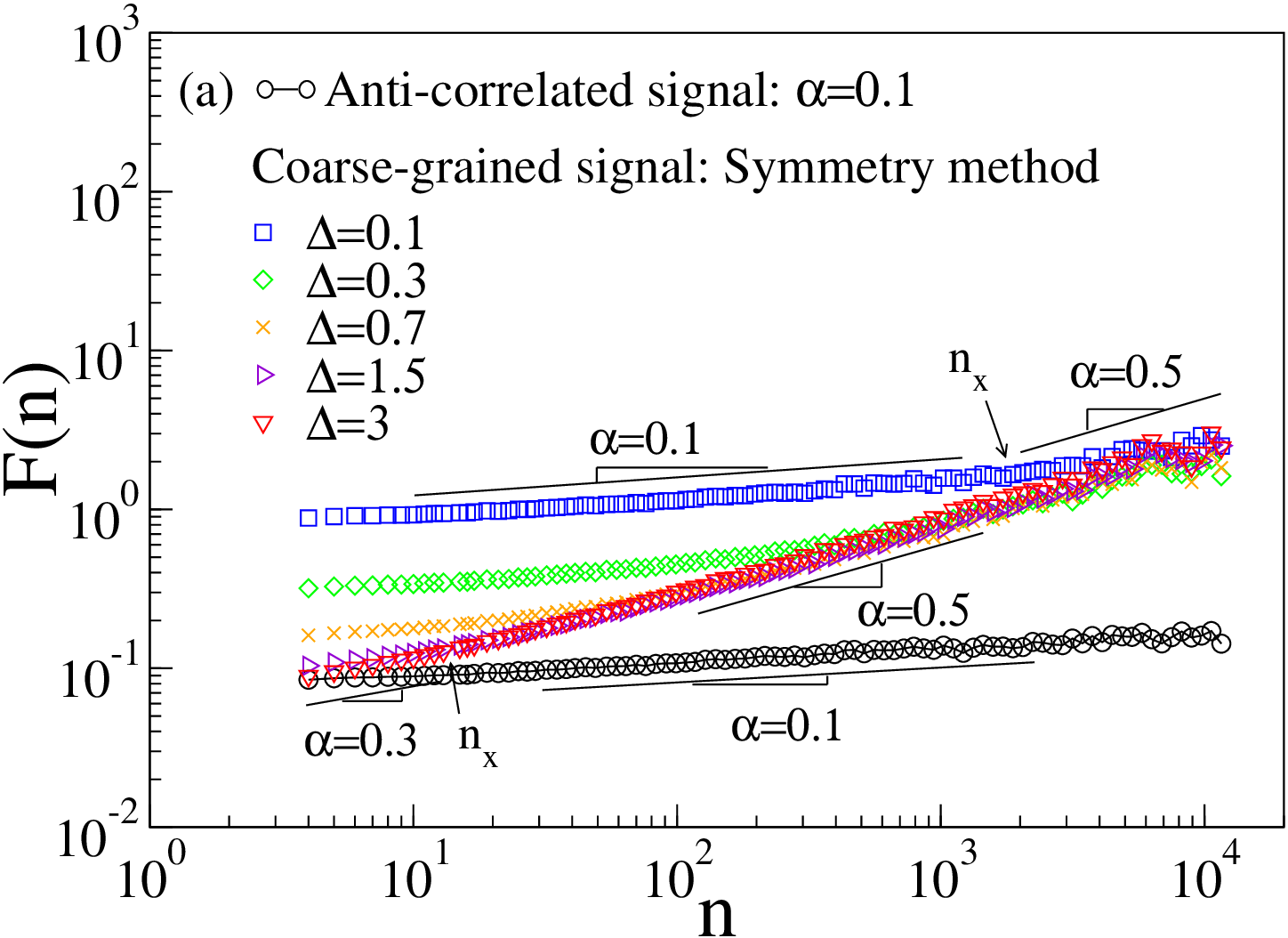}\vspace*{0cm}
    \includegraphics[width=0.85\columnwidth]{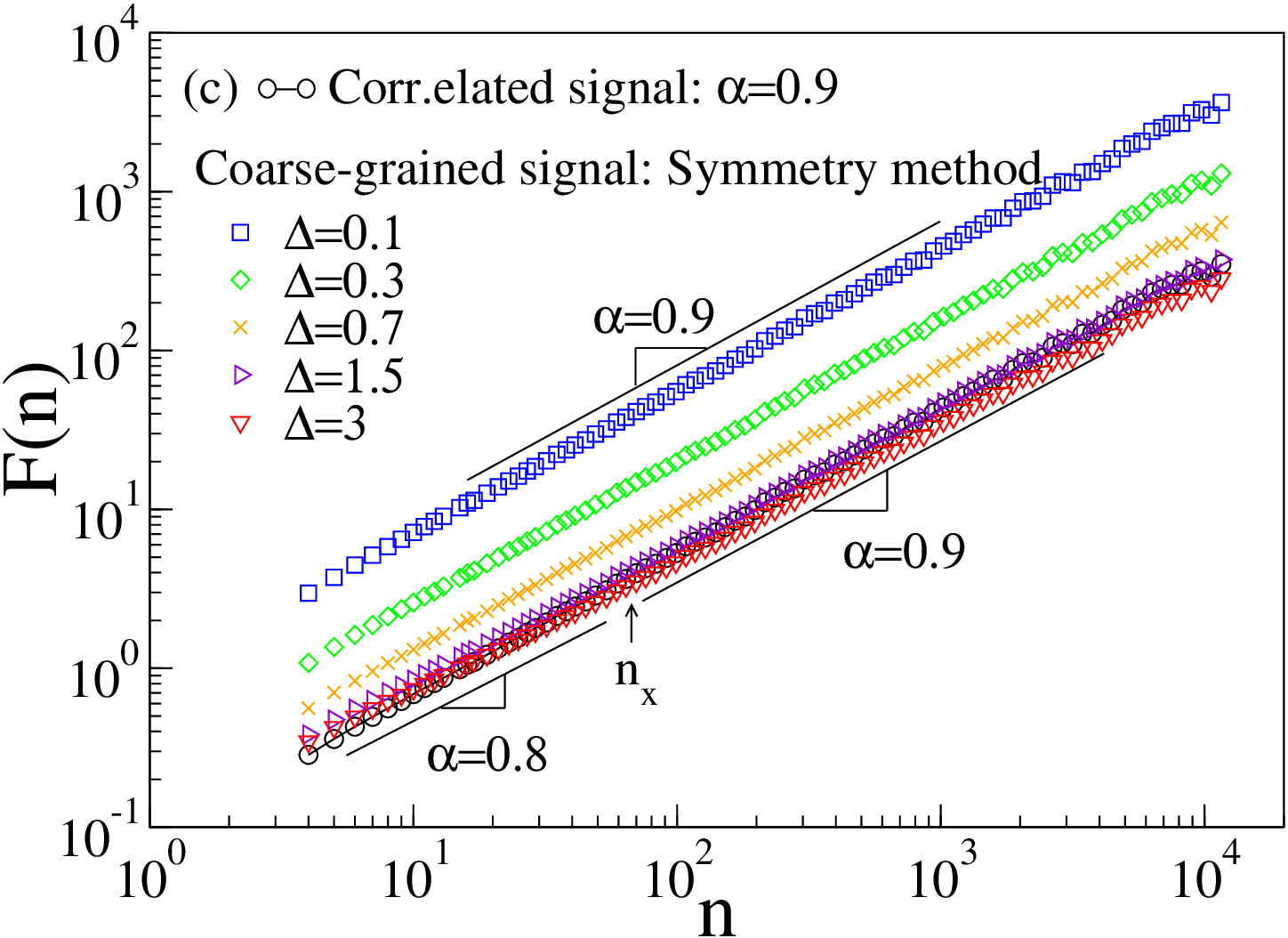}\vspace*{0cm}
    \includegraphics[width=0.85\columnwidth]{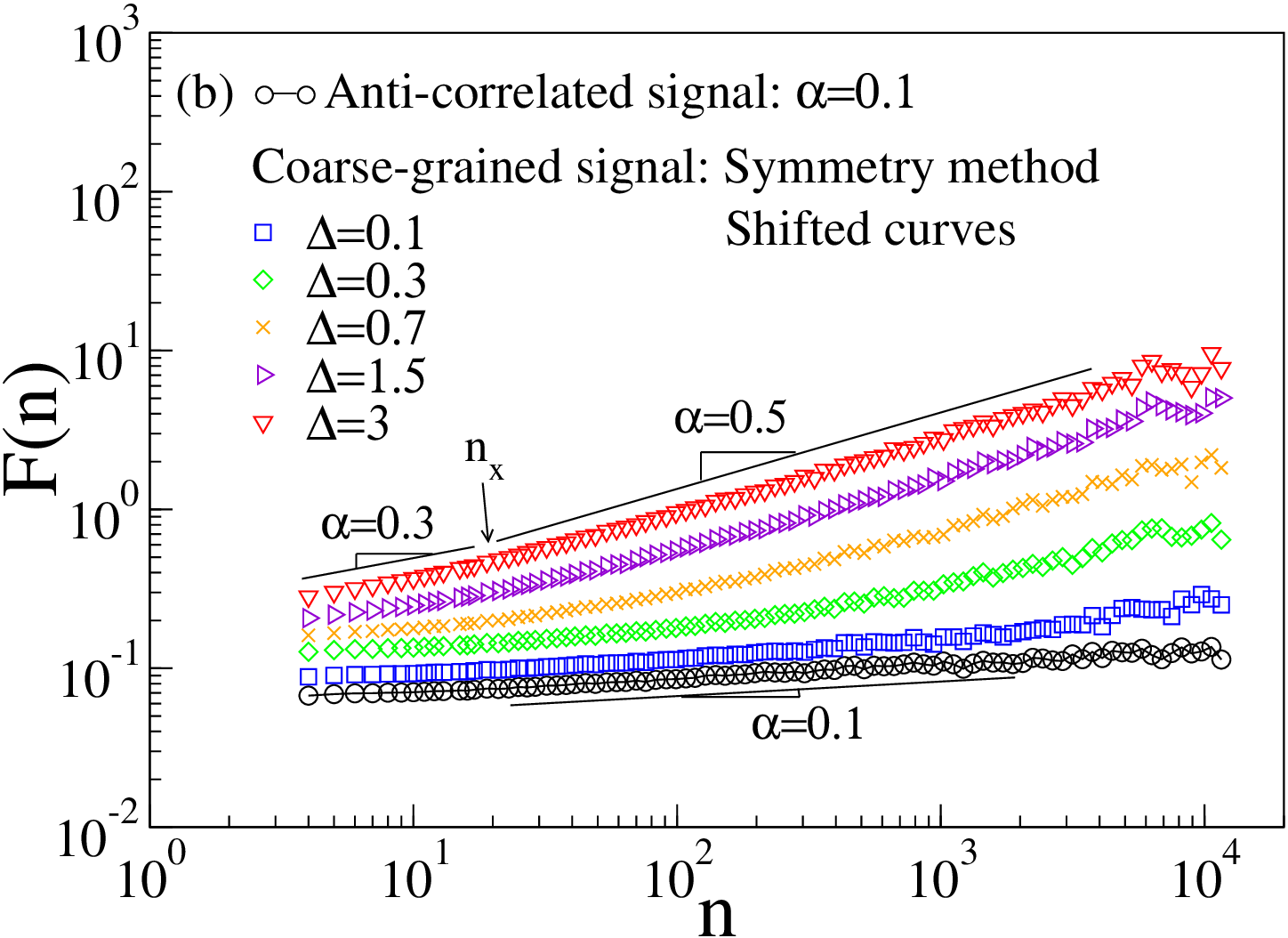}\vspace*{0cm}
    \includegraphics[width=0.85\columnwidth]{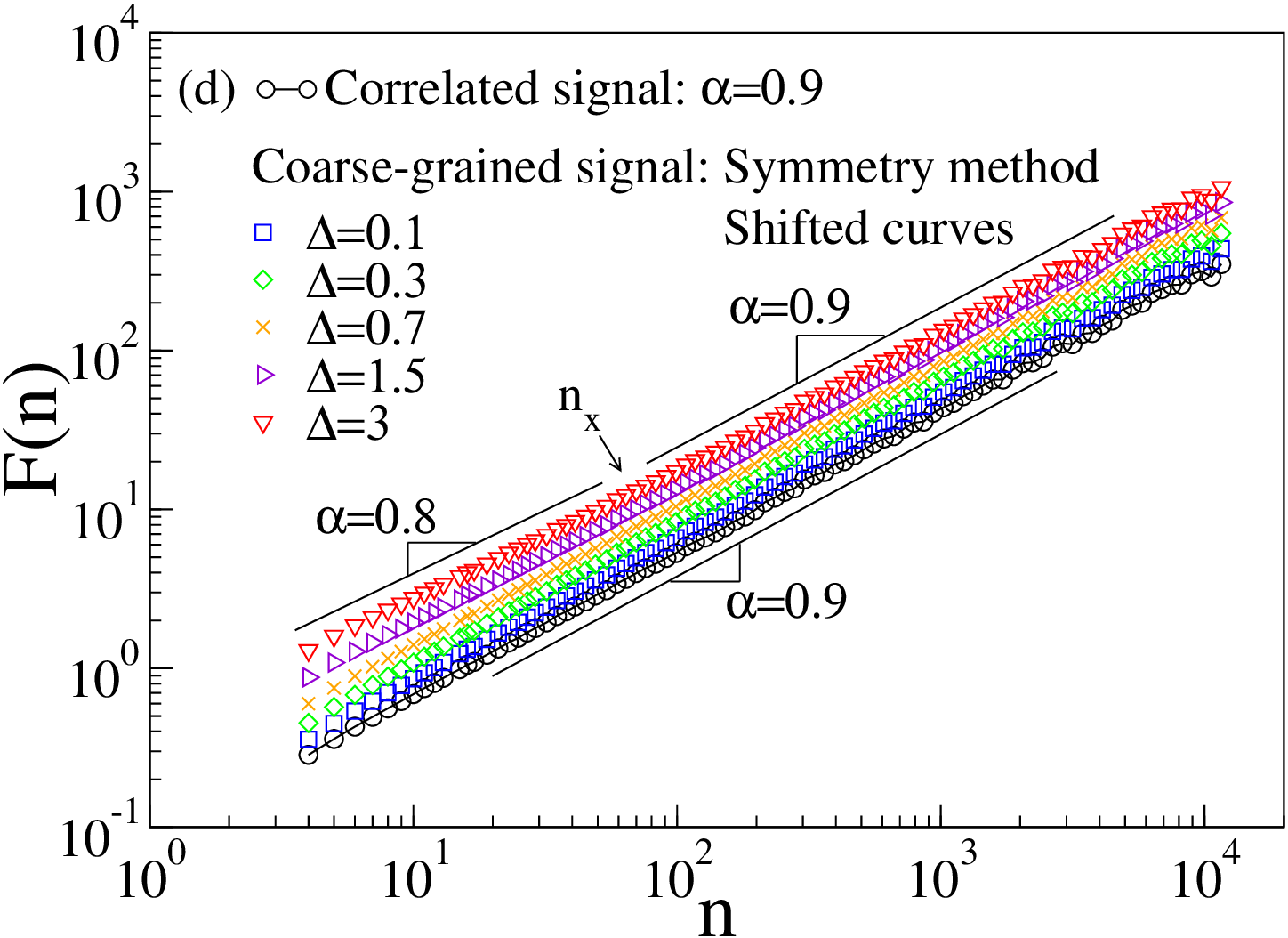}\vspace*{0cm}
    \includegraphics[width=0.85\columnwidth]{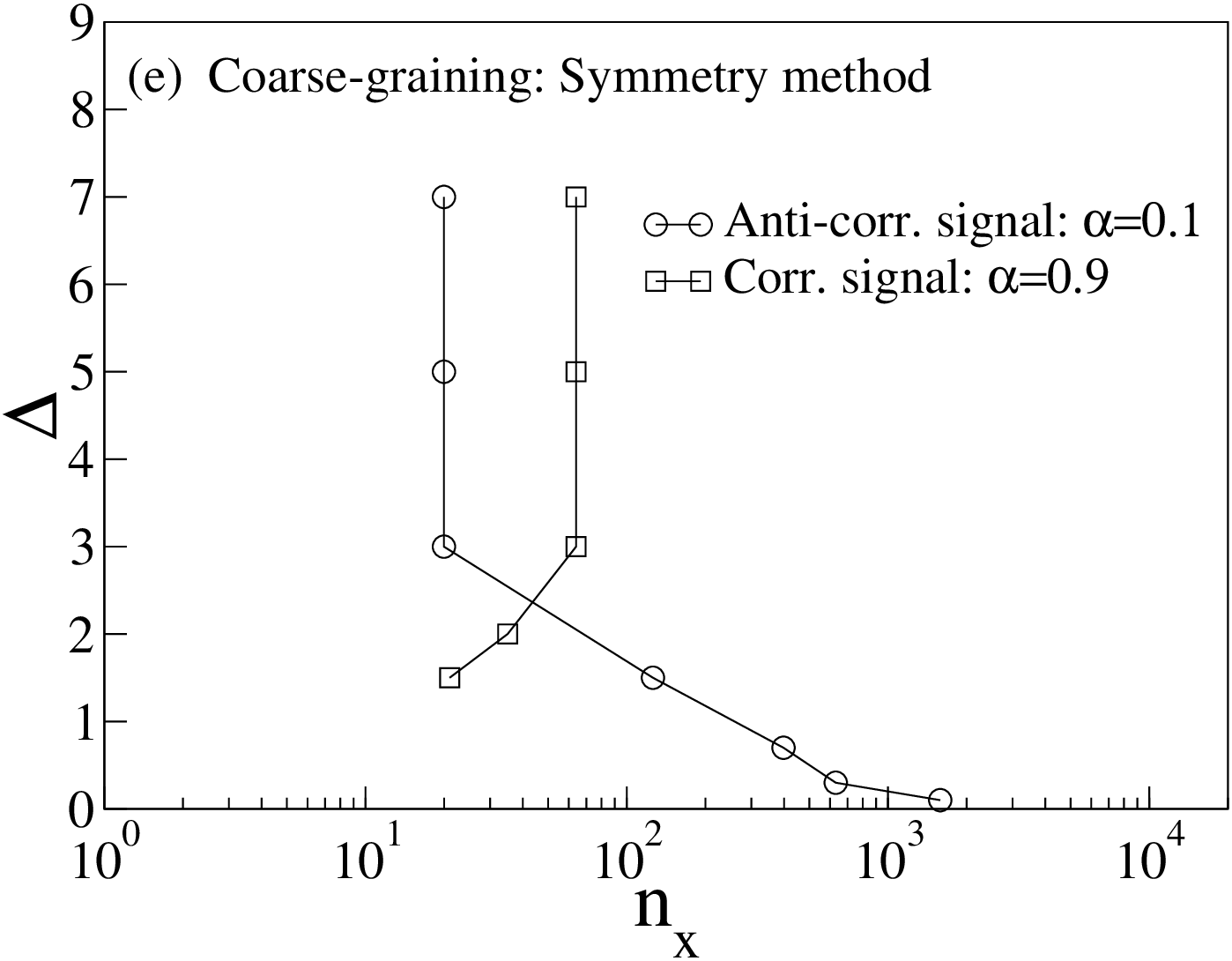}\vspace*{0cm}
    \caption { Effect of the {\it Symmetry coarse-graining method} on
      the scaling behavior of long-range power-law correlated signals.
      Scaling curves $F(n)$ obtained using the DFA-2 method of (a) an
      anti-correlated signal with scaling exponent $\alpha=0.1$, and
      (c)~a positively correlated signal with $\alpha=0.9$ before and
      after coarse-graining with different values of the partition
      parameter $\Delta$. The same signals as shown in
      Fig.\ref{fig:illustration of Symmetry coarse-grained signal}a, d
      are used for the analysis. For $\Delta<1$, the scaling curves
      $F(n)$ of the coarse-grained signals $\tilde{x}(i)$ shift up
      compared to $F(n)$ of the original signal $x(i)$ due to the
      expansion of the interface of $\tilde{x}(i)$ compared to $x(i)$
      (shown in Fig.\ref{fig:illustration of Symmetry coarse-grained
        signal}b, e). With increasing $\Delta>1$, the interface of
      $\tilde{x}(i)$ shrinks and $F(n)$ shifts down. Note, that for
      $\Delta>1$ the scaling curves $F(n)$ of the Symmetry
      coarse-grained signals remain above the $F(n)$ curves of the
      corresponding original signals, in contrast to the effect of the
      Floor method where $F(n)$ of the Floor coarse-grained signals
      shifts below $F(n)$ of the original signals. For the
      anti-correlated signal in (a), the Symmetry coarse-graining
      leads to a crossover from anti-correlated ($\alpha=0.1$) to
      random behavior ($\alpha=0.5$) at a crossover scale $n_x$. For
      $\Delta<1$ the crossover $n_x$ corresponds to large scales, and
      moves to intermediate and small scales with increasing
      $\Delta$. In contrast to anti-correlated signals, the Symmetry
      coarse-graining with small $\Delta<1$ has no effect on the
      scaling of positively correlated signals, while for large
      $\Delta>1$ there is a crossover at small scales to weaker
      correlations ($\alpha=0.8$). (b) and (d) show the same results
      as in (a) and (c) with scaling curves $F(n)$ vertically shifted
      to visualize the transition of the crossover $n_x$ for different
      values of $\Delta$. (e) Dependence of the crossover scale $n_x$
      on the coarse-graining parameter $\Delta$ for the
      anti-correlated and positively correlated signal shown in (a)
      and (c). For very rough Symmetry coarse-graining with $\Delta
      \ge 3$, the crossover scale $n_x$ does not change for both
      anti-correlated and positively correlated signals. The effect of
      the Symmetry coarse-graining method on the scaling of correlated
      and anti-correlated signals is very similar to that of the Floor
      method (compare with Fig.\ref{fig:Effect of Floor method
        coarse-graining on DFA}). }
\label{fig:Effect of Symmetric method coarse-graining on DFA}
\end{figure*}

{\it Correlated signals:} We find that the scaling behavior of
positively correlated signals is much less affected by the Floor
coarse-graining compared to anti-correlated signals. In
Fig.\ref{fig:Effect of Floor method coarse-graining on DFA}(c,d,e) we
present the results for a strongly correlated signal characterized by
a DFA scaling exponent $\alpha=0.9$. For small values of the width of
the partition interval $\Delta<1$ the interface of the coarse-grained
signal $\tilde{x}(i)$ expands along the $y$-axis, as demonstrated in
Fig.\ref{fig:illustration_of Floor coarse-grained signal}e, leading to
an increase and a vertical shift up of the DFA fluctuation function
$F(n)$ (Fig.\ref{fig:Effect of Floor method coarse-graining on
  DFA}c). Further, with increasing values of $\Delta>1$ the interface
of the coarse-grained correlated signal shrinks, i.e., the standard
deviation decreases (as shown in Fig.\ref{fig:illustration_of Floor
  coarse-grained signal}f), and the scaling curve $F(n)$ shifts
vertically down (Fig.\ref{fig:Effect of Floor method coarse-graining
  on DFA}c). This coarse-graining effect is very similar in both
positively correlated and anti-correlated signals (compare
Fig.\ref{fig:Effect of Floor method coarse-graining on DFA}a and
Fig.\ref{fig:Effect of Floor method coarse-graining on DFA}c).

In contrast to anti-correlated signals, the Floor coarse-graining with
small values of the width of the partition interval $\Delta<1$ does
not affect the scaling behavior of positively correlated signals and
the DFA exponent $\alpha$ remains unchanged at both small and large
scales $n$, i.e., no crossover in $F(n)$ is observed
(Fig.\ref{fig:Effect of Floor method coarse-graining on
  DFA}c,d). However, for values of the partition parameter $\Delta>1$,
the Floor coarse-graining leads to a crossover in the scaling curve
$F(n)$ to slightly weaker correlations (decreased scaling exponent
$\alpha$) at small scales $n$, whereas the scaling behavior at large
scales remains unchanged. This crossover is indicated by $n_x$ in
Fig.\ref{fig:Effect of Floor method coarse-graining on DFA}c. With
increasing $\Delta>1$, the crossover scale $n_x$ moves from small to
intermediate scales $n$. This crossover transition from small to
intermediate scales $n$ is shown in Fig.\ref{fig:Effect of Floor
  method coarse-graining on DFA}d where for clarity the scaling curves
$F(n)$ are vertically shifted up, relative to the original signal
$x(i)$ with $\alpha=0.9$, in the order of increasing values of
$\Delta$. We note, that even for large $\Delta \ge 3$, the scaling
exponent of the Floor coarse-grained signal $\tilde{x}(i)$ at small
scales decreases only slightly to $\alpha=0.8$ from the value
$\alpha=0.9$ for the original signal $x(i)$, indicating a much weaker
effect of the Floor coarse-graining on positively correlated signals
compared to anti-correlated signals.

For fine coarse-graining with $\Delta<1$ the relative disposition of
data values in the coarse-grained signal $\tilde{x}(i)$ is preserved
compared to the original correlated signal $x(i)$, as shown in
Fig.\ref{fig:illustration_of Floor coarse-grained signal}d and
Fig.\ref{fig:illustration_of Floor coarse-grained signal}e. As a
result, the scaling exponent $\alpha$ for $\tilde{x}(i)$ remains
unchanged --- the scaling curve $F(n)$ for $\Delta=0.1$, 0.3 and 0.7
remain parallel to $F(n)$ for the original signal $x(i)$ (see
Fig.\ref{fig:Effect of Floor method coarse-graining on DFA}d).
Increasing the width of the partition interval $\Delta>1$ leads to a
stronger coarse-graining effect at small and intermediate scales $n$
in contrast to anti-correlated signals where large and intermediate
scales are affected.

For very rough coarse-graining, with partition parameter $\Delta \ge
3$, we find that the crossover scale $n_x$ does not extend to large
scales $n$ and remains practically constant at $n_x \approx 60$,
and at scales $n>60$ the scaling exponent $\alpha$ is not affected
by the Floor coarse-graining (Fig.\ref{fig:Effect of Floor method
  coarse-graining on DFA}e). This is in contrast to anti-correlated
signals where Floor coarse-graining with $\Delta \ge 3$ affects all
scales (Fig.\ref{fig:Effect of Floor method coarse-graining on DFA}b
and Fig.\ref{fig:Effect of Floor method coarse-graining on DFA}e).  In
Fig.\ref{fig:Effect of Floor method coarse-graining on DFA}e, we
summarize our findings for the effect of the Floor coarse-graining on
the position of the crossover $n_x$ for both correlated and
anti-correlated signals. Specifically, we find that for
anti-correlated signals $n_x$ moves from large to small scales with
increasing $\Delta$, practically affecting the entire scaling. In
contrast, for positively correlated signals a crossover to weaker
correlations appears only for $\Delta>1$ at small scales $n_x$, while
the scaling behavior at intermediate and large scales remains
unaffected. We note that for very rough coarse-graining with $\Delta
\ge 3$ the crossover scale $n_x$ remains constant for both
anti-correlated and positively correlated signals.

The change we observe in the scaling behavior of power-law positively
correlated signals $x(i)$ after Floor coarse-graining is similar to
the effect of adding noise to each data point in $x(i)$ or of adding a
percentage of random spikes of a given amplitude to the original
signal. It was previously demonstrated that adding random spikes an
anti-correlated signal alters the scaling curve $F(n)$ leading to a
crossover to weaker correlations at small scales $n$ (see Fig.3e in
\cite{Zhichen_PRE_2002}). It was also reported that with increasing
the amplitude of the spikes and their concentration in the original
signal the crossover scale $n_x$ gradually moves from small to
intermediate scales \cite{Zhichen_PRE_2002}, an effect similar to the
effect of Floor coarse-graining we shown in Fig.\ref{fig:Effect of
  Floor method coarse-graining on DFA}(c,d). With increasing the
partition parameter $\Delta>1$ the difference between the values
$x(i)$ of the original correlated signal and the integer (symbolic)
values $\tilde{x}(i)$ in the coarse-grained signal also increases,
effectively equivalent to adding random noise.  Thus, as in the case
of anti-correlated signals, the effect of Floor coarse-graining on the
scaling of positively correlated signals is equivalent to the
superposition of the scaling curve $F(n)$ of the original signal
$x(i)$ with the scaling curve $F_{\eta}(n)$ of a white noise signal
($\alpha=0.5$). Increasing the amplitude and standard deviation of the
added white noise signal leads to an increase (a vertical shift up) in
$F_{\eta}(n)$, and thus, a shift of the intersection (i.e., the
crossover scale $n_x$) between $F_{\eta}(n)$ and $F(n)$ of the
original correlated signal from small to intermediate scales $n$ --- a
superposition rule first discussed in \cite{kun,Zhichen_PRE_2002}.
However, we note that the same superposition rule leads to
substantially different effects on the scaling of anti-correlated and
positively correlated signals --- while {large, intermediate and
  small} scales are {strongly} affected by increasing the
coarse-graining parameter $\Delta$ for anti-correlated signals, only
the {small} scales are {slightly} affected for positively correlated
signals.

(ii) {\it Symmetry Method}. Next, we consider how symbolic
coarse-graining in the magnitude based on the Symmetry method (see
Section \ref{Coarse-graining magnitude methods}, Eq.\ref{symmetry-1}
and Fig.\ref{fig:illustration_of Symmetry Method}) affects the scaling
behavior of long-range power-law correlated signals.  We find that the
scaling behavior of anti-correlated signals is much strongly affected
by the Symmetry coarse-graining compared to positively correlated
signals.

In Fig.\ref{fig:Effect of Symmetric method coarse-graining on DFA}a
and Fig.\ref{fig:Effect of Symmetric method coarse-graining on DFA}c
we present the results of Symmetry coarse-graining for a strongly
anti-correlated signal characterized by a DFA scaling exponent
$\alpha=0.1$ and a strongly correlated signal with
$\alpha=0.9$. Similarly to the Floor coarse-graining method, applying
the Symmetry coarse-graining method with small values of the width of
the partition interval $\Delta<1$, the interface of the coarse-grained
signal $\tilde{x}(i)$ expands along the $y$-axis
(Fig.\ref{fig:illustration of Symmetry coarse-grained signal}b and e)
leading to an increase in the standard deviation of the coarse-grained
signal, and to a vertical shift up of the DFA fluctuation function
$F(n)$ (Fig.\ref{fig:Effect of Symmetric method coarse-graining on
  DFA}a and c). With increasing values of $\Delta$ the interface of
the coarse-grained signal $\tilde{x}(i)$ shrinks, i.e., the standard
deviation decreases (as shown in Fig.\ref{fig:illustration of Symmetry
  coarse-grained signal}c and f), and the scaling curve $F(n)$ shifts
vertically down (Fig.\ref{fig:Effect of Symmetric method
  coarse-graining on DFA}a). This vertical shift of scaling curves
$F(n)$ is observed for both anti-correlated and positively correlated
Symmetry coarse-grained signals (Fig.\ref{fig:Effect of Floor method
  coarse-graining on DFA}a and c).

We note, that even for very large values of the Symmetry
coarse-graining parameter $\Delta>>1$ the scaling curves $F(n)$ of
both anti-correlated and positively correlated signals never shift
below the $F(n)$ curve of the original signal (Fig.\ref{fig:Effect of
  Symmetric method coarse-graining on DFA}a and c). This is in
contrast to the Floor coarse-graining method where for $\Delta>1$ the
scaling curve $F(n)$ of the coarse-grained signals (or at least a part
of $F(n)$ corresponding to small scales $n$) shift vertically down
below the scaling curve $F(n)$ of the original signal (see
Fig.\ref{fig:Effect of Floor method coarse-graining on DFA}a and c).
For $\Delta>>1$, both Floor and Symmetry coarse-graining method lead
to practically binary symbolic series $\tilde{x}(i)$. Since the values
of binary series $\tilde{x}_{\rm Floor}(i)$ produced by Floor
coarse-graining are either 0 or -1 (Fig.\ref{fig:illustration_of Floor
  coarse-grained signal}c and f), we can expect $\langle\tilde{x}_{\rm
  Floor}(i)\rangle\approx0.5$ and the standard deviation of the
coarse-grained signal $\sigma_{\rm Floor}\approx0.5$, which is smaller
than the standard deviation $\sigma=1$ of the original signal $x(i)$,
leading to a shift of the scaling curve $F(n)$ after the Floor
coarse-graining below $F(n)$ of the original signal $x(i)$. In
contrast, for Symmetry coarse-graining, the values of the binary
series are 1 and -1 (Fig.\ref{fig:illustration of Symmetry
  coarse-grained signal}c and f), leading to $\langle\tilde{x}_{\rm
  Symmetry}(i)\rangle\approx0$ and $\sigma_{\rm Symmetry}\approx1$,
and thus, the $F(n)$ curve of the Symmetry coarse-grained signal
$\tilde{x}(i)$ never shifts below the original $F(n)$ curve.

Coarse-graining signals using the Symmetry method leads to similar
crossover behaviors as we observed using the Floor method. We find
that the crossover effect after Symmetry coarse-graining is more
pronounced for anti-correlated signals compared to positively
correlated signals. Specifically, for anti-correlated signals there is
a crossover to a random behavior ($\alpha=0.5$) at a given scale
$n_x$, which with increasing the partition parameter $\Delta$ moves
from very large scales to intermediate and small scales (see
Fig.\ref{fig:Effect of Symmetric method coarse-graining on DFA}b where
the scaling curves $F(n)$ are vertically shifted for clarity). For
positively correlated signals, a crossover to slightly weaker
correlations at small scales appears only when applying the Symmetry
method with $\Delta>1$, and the crossover scale $n_x$ moves from small
to intermediate scales with increasing $\Delta$ (Fig.\ref{fig:Effect
  of Symmetric method coarse-graining on DFA}d).

The transitions of the crossover scale $n_x$ with increasing partition
parameter $\Delta$ using the Symmetry coarse-graining method are
illustrated for both anti-correlated and positively correlated
coarse-grained signal in Fig.\ref{fig:Effect of Symmetric method
  coarse-graining on DFA}e. Similar transition of the crossover scale
$n_x$ we also found for the Floor coarse-graining method
(Fig.\ref{fig:Effect of Floor method coarse-graining on DFA}e,
indicating that Symmetry and Floor coarse-graining method have a
similar effect on the scaling behavior of correlated signals. If we
ignore the few cases when data points $x(i)$ in the original signal
lie exactly on the boundaries of the partition intervals, the only
difference between the Symmetry and Floor coarse-graining methods is
that for values $x(i)>0$, $\tilde{x}_{\rm Symmetry}(i)=\tilde{x}_{\rm
  Floor}(i)+1$ (see Fig.\ref{fig:illustration_of Floor Method}b and
Fig.\ref{fig:illustration_of Symmetry Method}b). Thus, the relative
disposition of the data points $\tilde{x}(i)$ in the coarse-graining
signal is similar for the Symmetry or Floor method, leading to a very
similar effect on the scaling behavior of correlated signals (compare
Fig.\ref{fig:Effect of Floor method coarse-graining on DFA} and
Fig.\ref{fig:Effect of Symmetric method coarse-graining on DFA}).

\begin{figure*}
    \includegraphics[width=0.85\columnwidth]{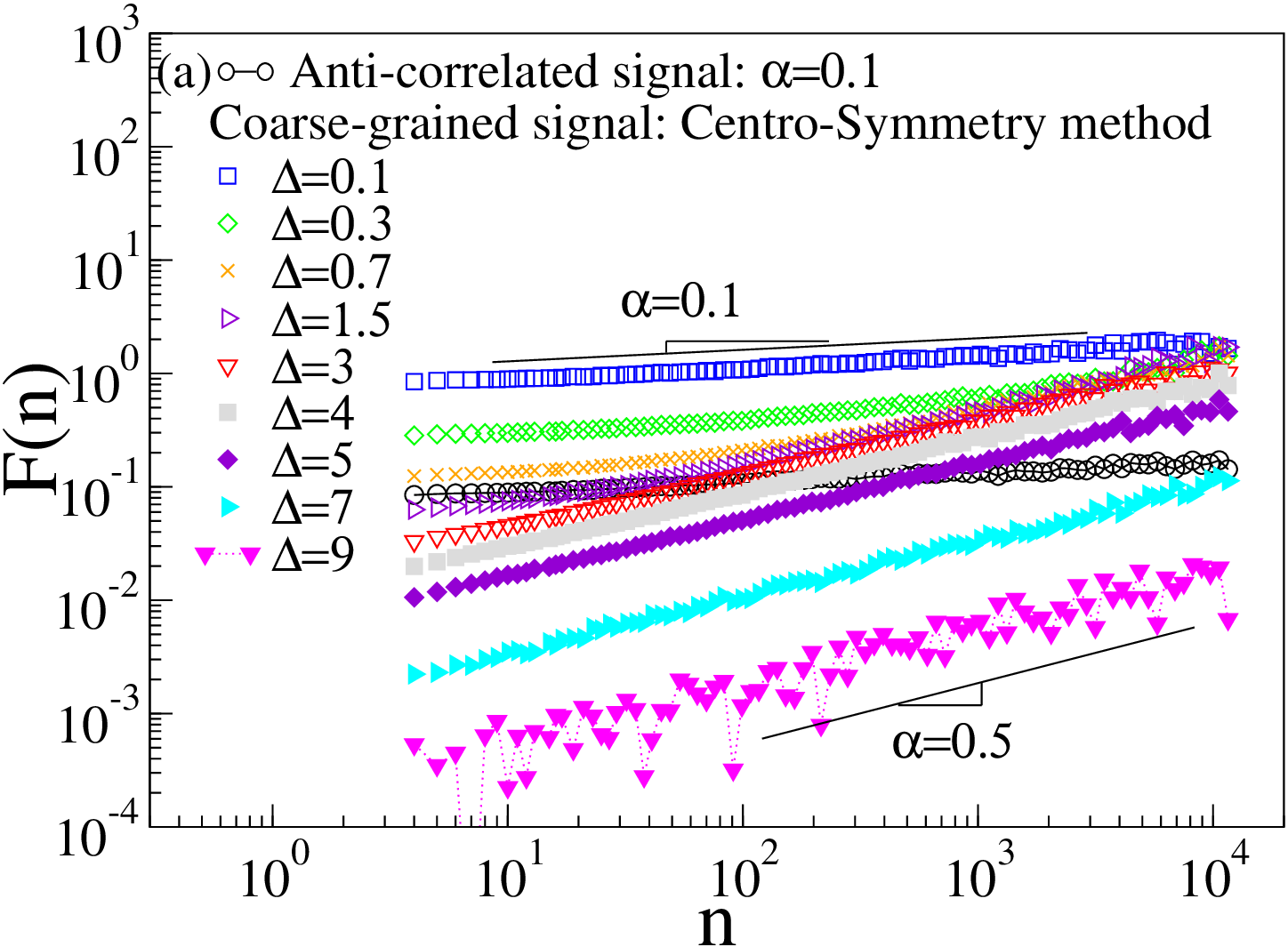}\vspace*{0cm}
    \includegraphics[width=0.85\columnwidth]{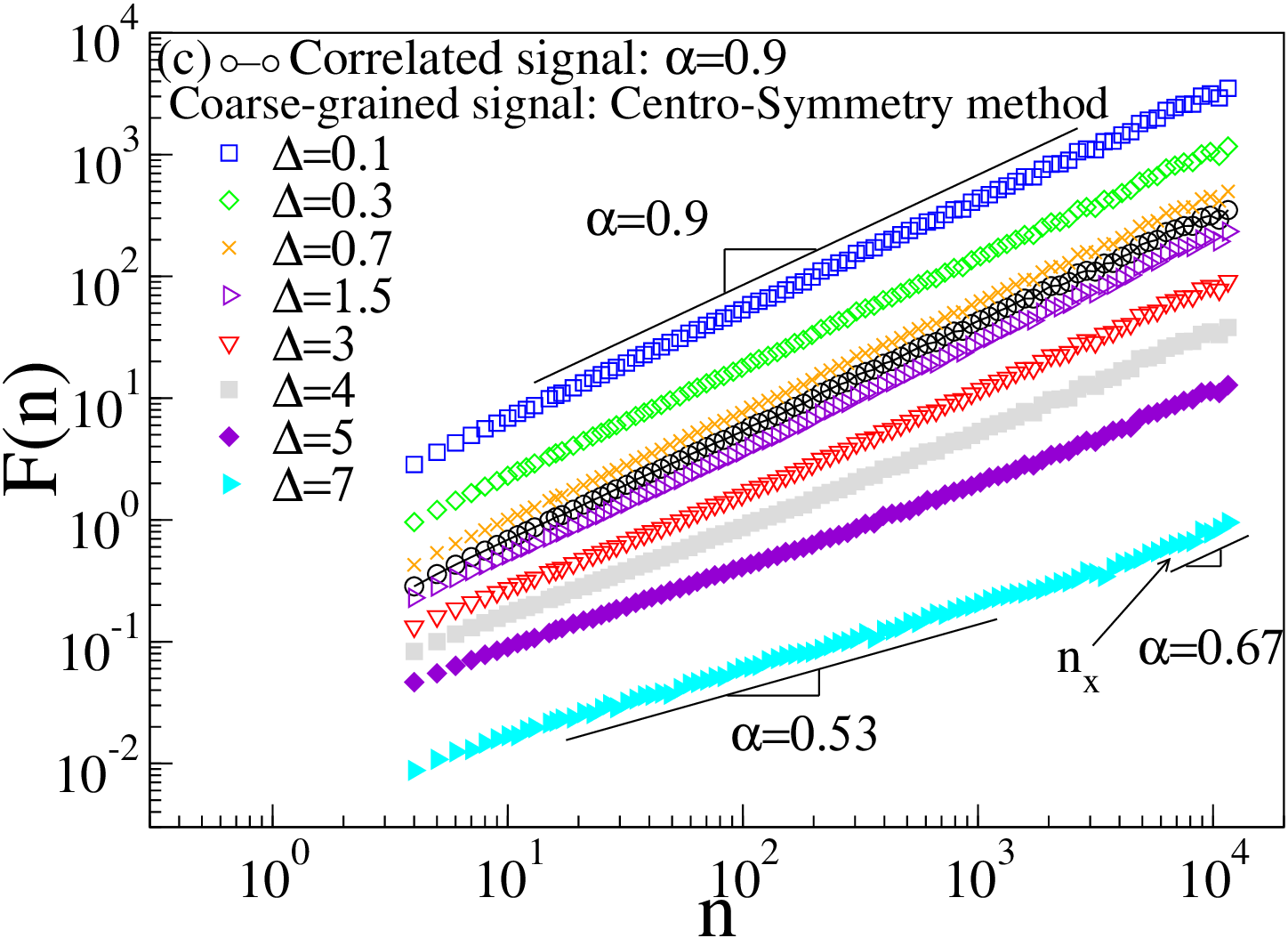}\vspace*{0cm}
    \includegraphics[width=0.85\columnwidth]{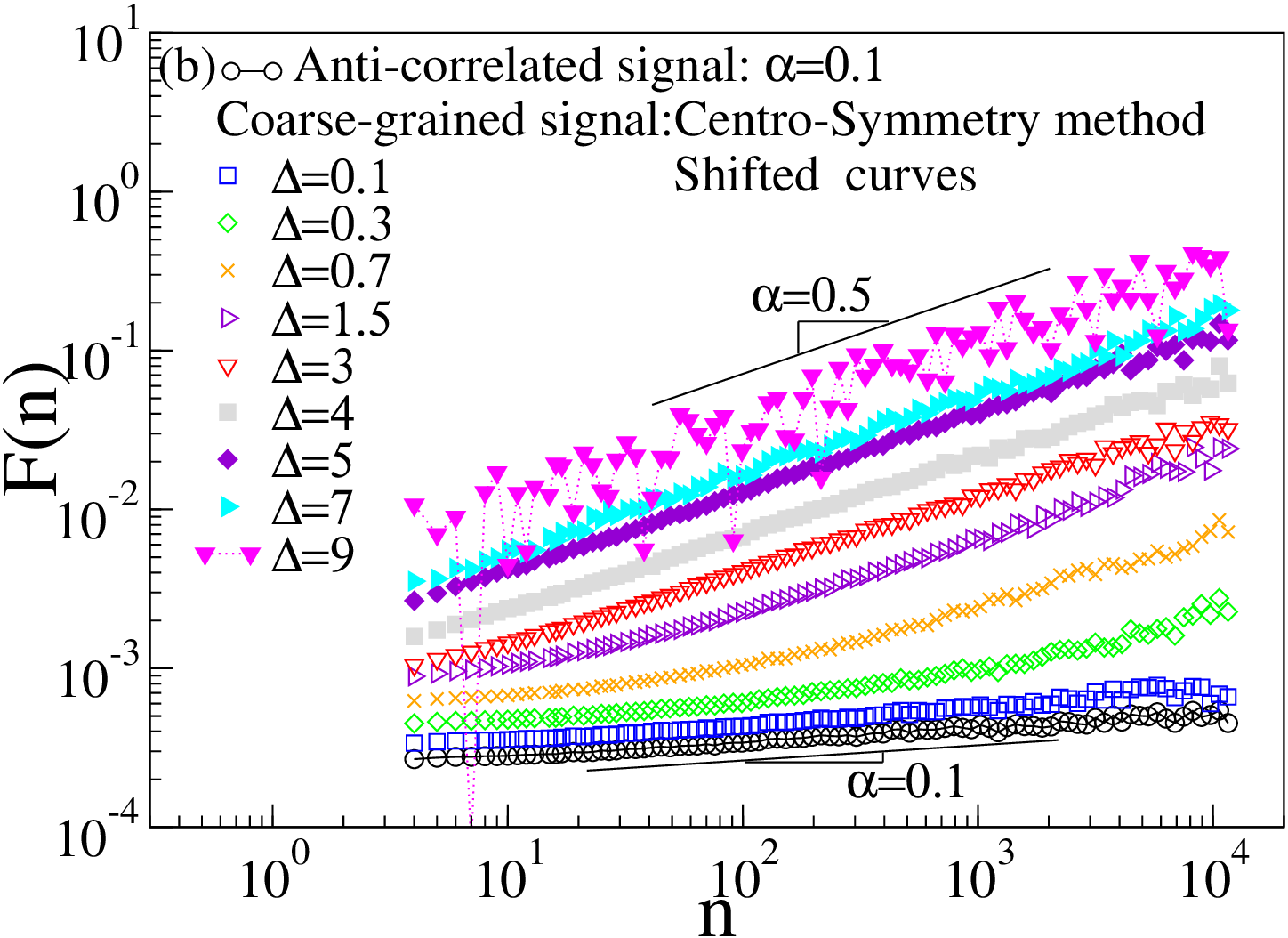}\vspace*{0cm}
    \includegraphics[width=0.85\columnwidth]{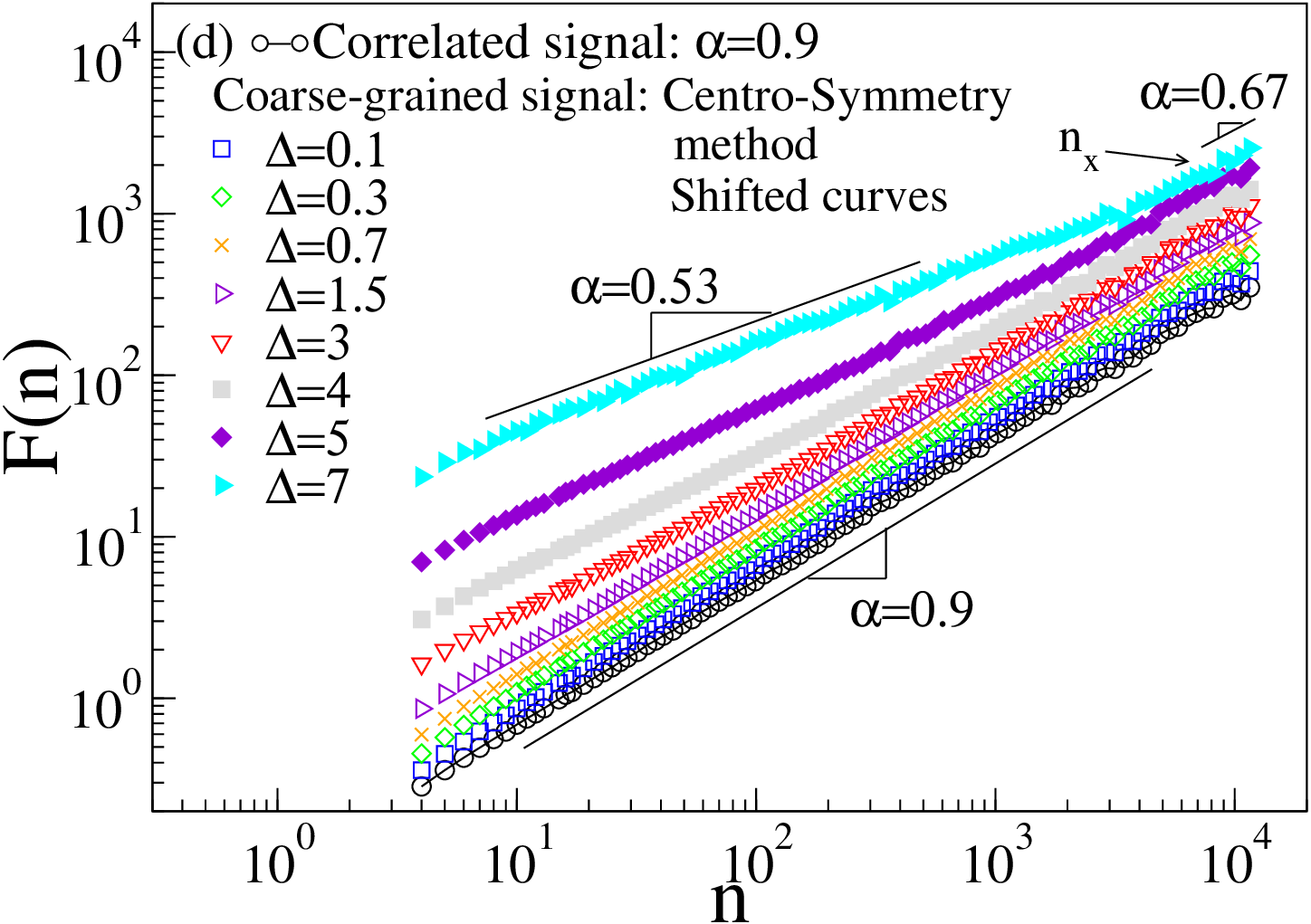}\vspace*{0cm}
    \includegraphics[width=0.85\columnwidth]{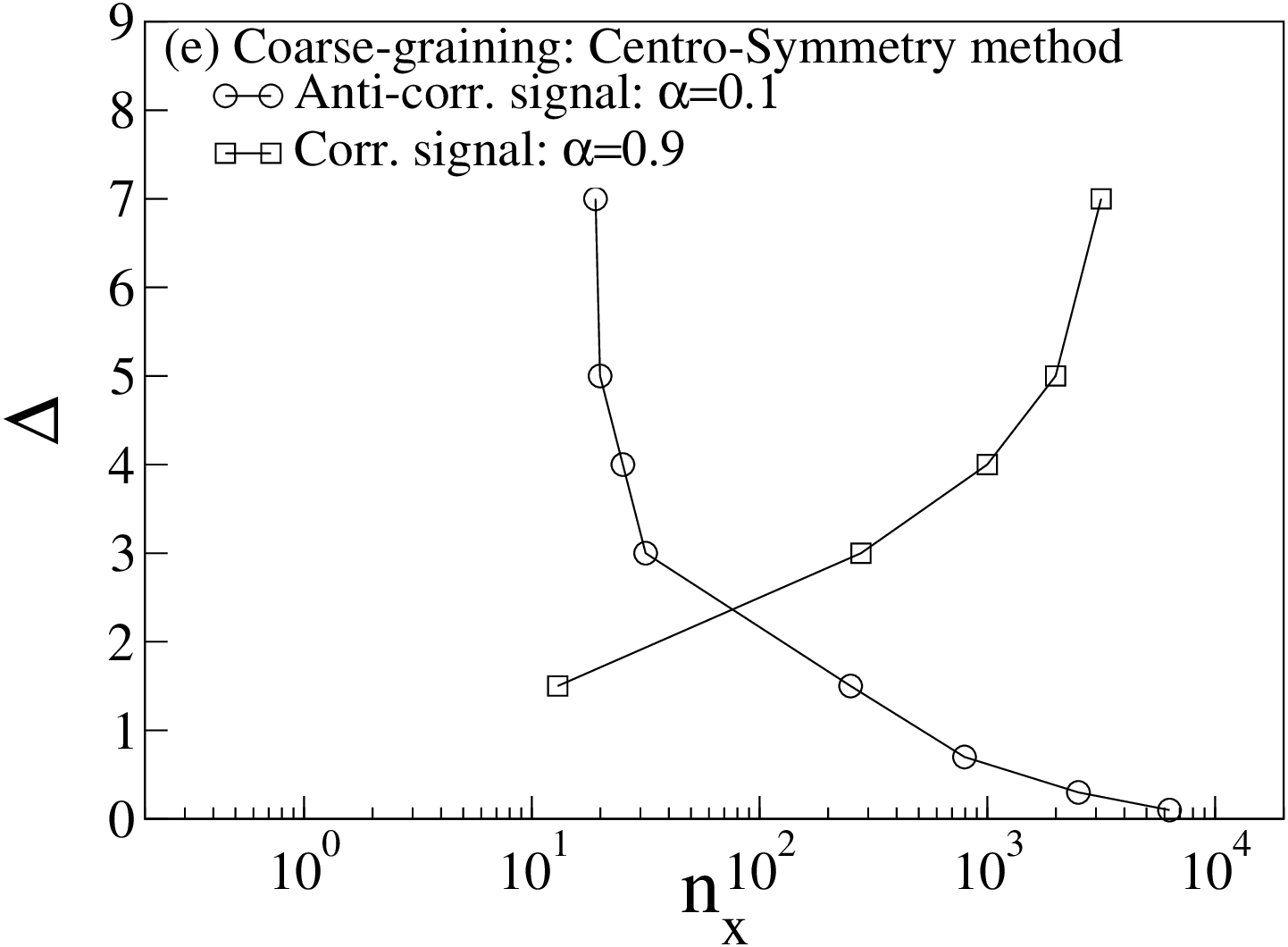}\vspace*{0cm}
    \caption { Effect of the {\it Centro-Symmetry coarse-graining
        method} on the scaling behavior of long-range power-law
      correlated signals. Scaling curves $F(n)$ obtained using the
      DFA-2 method of (a) an anti-correlated signal with scaling
      exponent $\alpha=0.1$, and (c) a positively correlated signal
      with $\alpha=0.9$ before and after coarse-graining with
      different values of the partition parameter $\Delta$. The same
      signals as shown in Fig.\ref{fig:illustration_of Centro-symmetry
        coarse-grained signal}a, d are used for the analysis. For
      $\Delta<1$, the scaling curves $F(n)$ of the coarse-grained
      signals $\tilde{x}(i)$ shift up compared to $F(n)$ of the
      original signal $x(i)$ due to the expansion of the interface of
      $\tilde{x}(i)$ compared to $x(i)$ (shown in
      Fig.\ref{fig:illustration_of Centro-symmetry coarse-grained
        signal}b, e). With increasing $\Delta>1$, the interface of
      $\tilde{x}(i)$ shrinks and $F(n)$ shifts down, similar to the
      Floor method (Fig.\ref{fig:Effect of Floor method
        coarse-graining on DFA}a, c) but different from the Symmetry
      method (Fig.\ref{fig:Effect of Symmetric method coarse-graining
        on DFA}a, c). For the anti-correlated signal in (a), the
      Centro-Symmetry coarse-graining leads to a crossover from
      anti-correlated ($\alpha=0.1$) to random behavior ($\alpha=0.5$)
      at a crossover scale $n_x$. For $\Delta<1$ the crossover $n_x$
      corresponds to large scales, and moves to intermediate and small
      scales with increasing $\Delta$. In contrast to anti-correlated
      signals, the Centro-Symmetry coarse-graining with small
      $\Delta<1$ has no effect on the scaling of positively correlated
      signals, while for large $\Delta>1$ there is a crossover at
      small scales towards random behavior ($\alpha \approx 0.5$) in
      contrast to the Floor and Symmetry methods. (b) and (d) show the
      same results as in (a) and (c) with scaling curves $F(n)$
      vertically shifted to visualize the transition of the crossover
      $n_x$ for different values of $\Delta$. (e) Dependence of the
      crossover scale $n_x$ on the coarse-graining parameter $\Delta$
      for the anti-correlated and positively correlated signals shown
      in (a) and (c). Note, that $n_x$ changes with increasing
      $\Delta>3$, in contrast to the Floor and Symmetry methods where
      $n_x$ stabilizes for $\Delta>3$ (Fig.\ref{fig:Effect of Floor
        method coarse-graining on DFA}e and Fig.\ref{fig:Effect of
        Symmetric method coarse-graining on DFA}e, see also the text
      for explanation). We note, that the effect of the
      Centro-Symmetry coarse-graining method is much stronger compared
      to the Floor and Symmetry methods (Fig.\ref{fig:Effect of Floor
        method coarse-graining on DFA} and Fig.\ref{fig:Effect of
        Symmetric method coarse-graining on DFA}).}

    \label{fig:Effect of Centro-Symmetric method coarse-graining on
      DFA}
\end{figure*}

The DFA scaling results we obtained here for anti-correlated signals
after rough coarse-graining with $\Delta\geqslant3$ are in agreement
with earlier studies of heartbeat interval increments, where after
applying the magnitude and sign decomposition method it was shown that
the sign ($+1$ or $-1$) of the heartbeat increments (i.e., a binary
sequence derived after "coarse-graining" of the original time series)
exhibits a scaling behavior with $\alpha\simeq0.3$ at small and
intermediate time scales with a crossover to $\alpha\simeq0.5$ at
large time scales $n>100$ \cite{Yosef2001,Yosef2003,Ivanov-PRE-2009}.

(iii) {\it Centro-Symmetry Method}. The last method of coarse-graining
in the magnitude we consider is the Centro-Symmetry method (see
Section \ref{Coarse-graining magnitude methods},
Eq.\ref{centro-symmetry-1} and Fig.\ref{fig:illustration of
  Centro-symmetry Method}). In Fig.\ref{fig:Effect of Centro-Symmetric
  method coarse-graining on DFA} we present the results for a strongly
anti-correlated signal characterized by a DFA scaling exponent
$\alpha=0.1$ and for a positively correlated signal with $\alpha=0.9$
(the same identical signals are used for comparison with the Floor and
Symmetry coarse-graining in Fig.\ref{fig:Effect of Floor method
  coarse-graining on DFA} and Fig.\ref{fig:Effect of Symmetric method
  coarse-graining on DFA} respectively). We observe a vertical shift
of the DFA fluctuation function $F(n)$ after Centro-Symmetry
coarse-graining of both anti-correlated and positively correlated
signals (Fig.\ref{fig:Effect of Centro-Symmetric method
  coarse-graining on DFA}a and c), i.e., a shift up for values of the
partition parameter $\Delta<1$ and a shift down for $\Delta>1$. This
effect is very similar to the Floor coarse-graining
(Fig.\ref{fig:Effect of Floor method coarse-graining on DFA}a and c),

As for the Floor and Symmetry coarse-graining, for small values of
$\Delta$ the Centro-Symmetry method applied to anti-correlated signal
leads to a crossover at large scales to random behavior with an
exponent $\alpha=0.5$ (Fig.\ref{fig:Effect of Centro-Symmetric method
  coarse-graining on DFA}b). With increasing $\Delta$ the crossover
scale $n_x$ moves to intermediate and small scales. Applied to
positively correlated signals the Centro-Symmetry coarse-graining with
partition parameter $\Delta<1$ does not lead to observable changes in
the scaling behavior and the scaling exponent $\alpha=0.9$ remains
unchanged across all scales $n$ (Fig.\ref{fig:Effect of
  Centro-Symmetric method coarse-graining on DFA}c and d) --- similar
to the Floor and Symmetry methods. In contrast, for increasing
$\Delta>1$ the Centro-Symmetry coarse-graining leads to a dramatic
change from strongly correlated ($\alpha=0.9$) towards uncorrelated
($\alpha=0.5$) behavior with a crossover scale $n_x$ moving from small
to intermediate and large scales (Fig.\ref{fig:Effect of
  Centro-Symmetric method coarse-graining on DFA}d). This effect of
the Centro-Symmetry method is much stronger compared to the Floor and
Symmetry methods, where for positively correlated signals the
coarse-graining leads to slightly weaker correlations ($\alpha=0.8$)
only at small scales (Fig.\ref{fig:Effect of Floor method
  coarse-graining on DFA}d and Fig.\ref{fig:Effect of Symmetric method
  coarse-graining on DFA}d).

The transition of the crossover scale $n_x$ for increasing partition
parameter $\Delta$ of the Centro-Symmetry coarse-graining of
correlated signals is shown in Fig.\ref{fig:Effect of Centro-Symmetric
  method coarse-graining on DFA}e. For anti-correlated signals the
crossover scale $n_x$ moves from large to small scales with increasing
$\Delta$. We note that, in contrast to the Floor and Symmetry methods,
$n_x$ continues to decrease to smaller scales even when
$\Delta>3$. For correlated signals $n_x$ moves in the opposite
direction, from small to large scales with increasing $\Delta>1$,
which is in contrast to the Floor and Symmetry coarse-graining where
$n_x$ does not reach to large scales and remains stable for $\Delta>3$
(Fig.\ref{fig:Effect of Floor method coarse-graining on DFA}e and
Fig.\ref{fig:Effect of Symmetric method coarse-graining on DFA}e).
The reason for this stable behavior of $n_x$ is that for increasing
$\Delta>3$ most of the data points $\tilde{x}(i)$ become 1 and -1 for
the Symmetry method and 0 and -1 for the Floor method with standard
deviation of the coarse-grained signals $\sigma_{\rm
  Symmetry}=1+\varepsilon$ and $\sigma_{\rm Floor}=0.5+\varepsilon$,
where the term $\varepsilon \ge 0$ is due to a few outliers $\left|
  x(i) \right| > \Delta$ and $\varepsilon \to 0$ for large
$\Delta$. For $\Delta>3$ the relative contribution of $\varepsilon$ to
$\sigma_{\rm Symmetry}$ and $\sigma_{\rm Floor}$ is negligible,
$\sigma_{\rm Symmetry}$ and $\sigma_{\rm Floor}$ stabilize, and thus
there is no further change in the scaling curves $F(n)$ and the
crossover scale $n_x$. In contrast, most of the data points after
Centro-Symmetry coarse-graining become $\tilde{x}(i)=0$, with a
standard deviation $\sigma_{\rm Centro-Symmetry}=\varepsilon$, where
the term $\varepsilon$ is due to a few remaining outliers $\left| x(i)
\right| > \frac{1}{2}\Delta$. Because $\varepsilon \to 0$ with
increasing $\Delta>3$, $\sigma_{\rm Centro-Symmetry}$ also
continuously decreases and does not stabilize, and thus the scaling
curve $F(n)$ and the crossover scale $n_x$ continue to change as we
observe in Fig.\ref{fig:Effect of Centro-Symmetric method
  coarse-graining on DFA}.

\subsection{Coarse-graining in time}

\begin{figure}
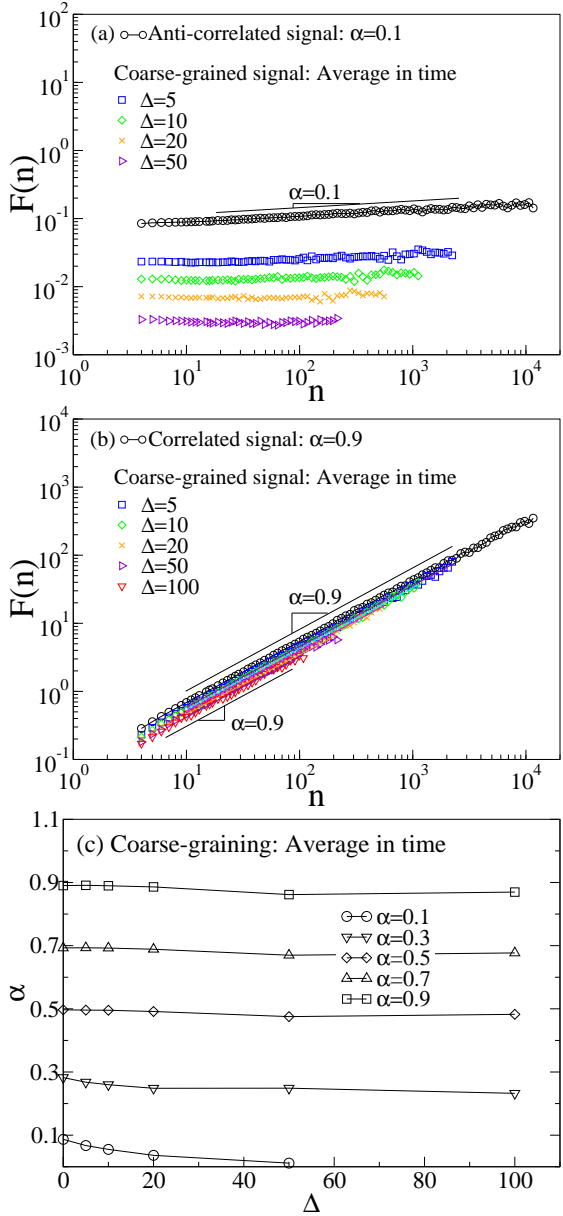

    \includegraphics[width=0.85\columnwidth]{Fig13a}\vspace*{0cm}
    \includegraphics[width=0.85\columnwidth]{Fig13b}\vspace*{0cm}
    \includegraphics[width=0.85\columnwidth]{Fig13c}\vspace*{0cm}
    \caption { Effect of {\it coarse-graining in time} on the scaling
      behavior of long-range power-law correlated signals.  Scaling
      curves $F(n)$ obtained using the DFA-2 method of (a) an
      anti-correlated signal with scaling exponent $\alpha=0.1$, and
      (c) a positively correlated signal with $\alpha=0.9$ before and
      after coarse-graining with different values of the length of the
      averaging time interval $\Delta$. The same signals as shown in
      Fig.\ref{fig:Example of coarse-graining in time}a,e are used for
      the analysis. The scaling curves $F(n)$ of the coarse-grained
      signals $\tilde{x}(i)$ are shortened due to the coarse-graining
      in time, and are shifted down with increasing $\Delta$ due to
      decreasing of the standard deviation of $\tilde{x}(i)$ compared
      to the standard deviation of the original signal $x(i)$. (c)
      Dependence of the scaling exponent $\alpha$ obtained for the
      coarse-grained signals on the size of the averaging time
      interval $\Delta$. The scaling behavior of the coarse-grained
      signals is practically preserved for both anti-correlated and
      positively correlated signals. A slight decrease in $\alpha$ for
      anti-correlated signals is observed. }

    \label{fig:Effect of Average in time method coarse-graining on
      DFA}
\end{figure}

In addition to coarse-graining the magnitude of a signal, we also
consider how coarse-graining in time (see Section \ref{Coarse-graining
  time method}, Eq.\ref{coarse-graining in time} and
Fig.\ref{fig:Coarse-graining in time}) affects the scaling behavior of
long-range power-law correlated signals. We systematically investigate
the effect of time coarse-graining on both anti-correlated and
positively correlated signals. In Fig.\ref{fig:Effect of Average in
  time method coarse-graining on DFA}a we present the results for a
strongly anti-correlated signal with scaling exponent $\alpha=0.1$. We
find that after averaging in time the standard deviation of the signal
decreases, leading to a vertical shift down of the DFA fluctuation
function $F(n)$ for increasing averaging time intervals
$\Delta$. Although the averaging procedure in non-overlapping
intervals $\Delta$ substantially shortens the signal (up to 50 times),
the correlation properties remain practically preserved. For positively
correlated signals (Fig.\ref{fig:Effect of Average in time method
  coarse-graining on DFA}b), coarse-graining also causes shift down of
the scaling curve $F(n)$, however to a less extent compared to
anti-correlated signals. The scaling of the positively correlated
signals is well preserved after coarse-graining in time. In
Fig.\ref{fig:Effect of Average in time method coarse-graining on DFA}c
we present the dependence of scaling exponent $\alpha$ on the
coarse-graining time intervals $\Delta$ for various correlated
signals. We find that for anti-correlated signals $\alpha$ slightly
decreases after coarse-graining, and that this decrease is more
pronounced for larger $\Delta$ and for signals with stronger
anti-correlations. In contrast, the scaling exponent of positively
correlated signals remains practically unchanged after coarse-graining
with time intervals $\Delta$ up to 100.

\section{Summary}\label{Summary}

In summary, we systematically study the effect of coarse-graining in
both the magnitude and time on the scaling behavior of long-range
power-law anti-correlated and correlated signals. We investigate three
types of coarse-graining methods in the magnitude --- the Floor, the
Symmetry and the Centro-Symmetry method. For coarse-graining in time,
we employ a time average procedure within non-overlapping windows of
fixed size.

We find that, for both anti-correlated and positively correlated
signals, all three types of coarse-graining in the magnitude lead to
expanding the interface of the coarse-grained signal along the
$y$-axis for small values of the width of the partition interval
$\Delta<1$, and to a corresponding increase and a vertical shift up of
the DFA fluctuation function $F(n)$.  For increasing values of
$\Delta$, the interface of the coarse-grained signals shrink and the
scaling curves $F(n)$ shift vertically down.

For anti-correlated signals, all three coarse-graining methods in the
magnitude have a strong and similar effect on the scaling behavior of
the coarse-grained signals, leading to a crossover at scale $n_x$ to
random behavior ($\alpha=0.5$) at large scales.  With increasing
values of partition parameter $\Delta$, the crossover scale $n_x$
moves from large to intermediate and small scales.  In contrast, all
three coarse-graining methods have a much weaker effect on the scaling
behavior of positively correlated signals.  The scaling behavior of
coarse-grained positively correlated signals with partition parameter
$\Delta<1$ remains practically unchanged at both small and large
scales, i.e., there is no observable crossover in the scaling curve
$F(n)$.  For large $\Delta$, a crossover to slightly weaker
correlations at small scales appears after coarse-graining positively
correlated signals, and gradually moves to intermediate scales with
increasing $\Delta$.

For very large $\Delta>3$, the effect of Floor and Symmetry
coarse-graining on the scaling behavior and the position of the
crossover stabilizes for both anti-correlated and positively
correlated signals.  In contrast, the effect of Centro-Symmetry
coarse-graining on the scaling behavior is continuously stronger with
increasing $\Delta>3$, and the scale of the crossover $n_x$ is
continuously moving.  For Centro-Symmetry coarse-graining with very
large $\Delta>>3$, the correlations of strongly correlated signals are
destroyed ($\alpha \approx 0.5$) at small scales, and the correlations
at large scales are weakened.

The change we observe in the scaling behavior of both anti-correlated
and positively correlated signals $x(i)$ after coarse-graining in the
magnitude is to some extent similar to the effect of adding noise or
random spikes to long-range power-law correlated signals.  With
increasing the partition parameter $\Delta$ the difference between the
increments of the consecutive values $x(i)$ in the original signal and
the increments of the integer (symbolic) values $\tilde{x}(i)$ in the
coarse-grained signal also increases, effectively equivalent to adding
random noise.  It was previously shown \cite{Zhichen_PRE_2002} that
adding noise and random spikes to anti-correlated signals leads to a
crossover to a random behavior at large and intermediate scales, while
for positively correlated signals, a crossover to random behavior
occurs at small and intermediate scales --- a change in the scaling
behavior which strongly resembles the effects of coarse-graining
investigated here, and which can be explained by a superposition of
the scaling curve $F(n)$ of the original signal and the scaling curve
$F_{\eta}(n)$ of a white noise signal intersecting at a crossover
scale $n_x$~\cite{kun,Zhichen_PRE_2002}.

For coarse-graining in time where data are averaged in non-overlapping
time windows, we find that the scaling curve $F(n)$ shifts down due to
the decrease in the standard deviation of the coarse-grained signal,
and that the scaling is practically preserved for positively
correlated signals. While for anti-correlated signals there is a
slight decrease in the scaling exponent.

The reported here observations are important to correctly interpret
results of correlation and scaling analyses of a broad range of
natural and pre-processed symbolic sequences and coarse-grained
signals, and are instrumental in supporting modeling efforts of
systems generating symbolic dynamics.

\section*{Acknowledgments}
We thank the Brigham and Women's Hospital Biomedical Research
Institute Fund, the Spanish Junta de Andalucia (Grant
No. P06-FQM1858 and P07-FQM3163),
the Ministry of Human Resources of P.R.China (Scientific Research
Foundation for the Returned Overseas Chinese Scholars), Nanjing
Normal University (Scientific Research Foundation for the Returned
Overseas Chinese Scholars),
and the National Natural Science Foundation of China (Grant
No. 60501003) for support.

\end{document}